\documentclass[a4paper,11pt]{article}
\usepackage{latexsym}
\usepackage{feynmf}		
\usepackage{graphicx}
\def\ba{\begin{eqnarray}}
\def\ea{\end{eqnarray}}

\def\lb{\label}
\def\be{\begin{equation}}
\def\ee{\end{equation}}

\def\theequation{\arabic{section}.\arabic{equation}}

\begin{document}
\baselineskip0.25in
\title{Aspects of electrostatics in  BTZ geometries}
\author{Y. Herrera $^{1}$\thanks{e-mail: yagogaoc@hotmail.com}\, V. Hurovich $^{1}$ \thanks{e-mail: valehurovich@gmail.com}O. Santill\'an $^{2}$\thanks{e-mail: firenzecita@hotmail.com, osantil@dm.uba.ar}\, and C. Simeone $^{1}$\thanks{e-mail: csimeone@df.uba.ar}\, \,.\date {}
\\
\\ \\
{\small $^1$ Departamento de F\'isica, Facultad de Ciencias Exactas y Naturales,}\\
{\small Universidad de Buenos Aires and IFIBA, CONICET, Cuidad Universitaria, Buenos Aires 1428, Argentina.}\\
\\
{\small $^2$ CONICET--Instituto de Investigaciones Matem\'aticas Luis Santal\'o,} \\
{\small Ciudad Universitaria Pab. I, Buenos Aires 1428, Argentina.} }
\maketitle

\begin{abstract}
In the present paper  the electrostatic of charges in non rotating BTZ black hole  and wormhole space times is studied.  In particular, the self force of  a point charge in the geometry is characterized analitically. The differences between the self force in both cases is a theoretical experiment for distinguishing both geometries, which otherwise are locally indistinguishable. This idea was applied before to four and higher dimensional black holes by the present and other authors.  However, the particularities of the BTZ geometry makes the analysis considerable more complicated than usual electrostatic
in a flat geometry, and its even harder than its four dimensional counterparts. First, the BTZ space times are not asymptotically flat but instead asymptotically AdS. In addition, the relative distance $d(r,r+1)$ between two particles located at a radius $r$ and $r+1$ in the geometry tends to zero when $r\to\infty$. This behavior, which is radically different in a flat geometry, changes the analysis of the asymptotic conditions for the electrostatic field. In addition, there are no summation formulas that allow a closed analytic expression for the self force. We find a method to calculate such force in series, and the resulting expansion  is convergent to the real solution. However, we suspect that the convergence is not uniform. In other, for points that are far away from the black hole  the calculation of the force requires higher order summation. These subtleties are carefully analyzed in the paper, and it is shown that they lead to severe problems when calculating the self force for asymptotic points in the geometry. 
 \end{abstract}
\section{Introduction}

Electrodynamics in General Relativity is described by the Maxwell equations in curved space-time \cite{ll}. A freely falling observer in such background would write the same equations valid for Minkowski space-time; however, these equations must have a different solution, because the curved geometry imposes a different asymptotic behavior than the flat one. In particular, the electric field around a static point charge in a curved background is not spherically symmetric in general, and this gives a non zero electrostatic self-force on the charge.

One of the earliest studies on the electrostatic self-force on static charges induced by a curved background was that on a Schwarzschild black hole geometry \cite{will80}. In that reference it was shown that the self-force on a charge $q$ is repulsive, i.e. it points outwards from the black hole, and the functional dependence on the position is given by
$$
f\sim\frac{mq^2}{r^3},
$$
where $2m$ is the horizon radius of the black hole and $r$ is the Schwarzschild radial coordinate of the charge. This result was first obtained within the framework of linearized general relativity \cite{vil79}, and was later recovered working within the full theory \cite{will80}.

 After the publication of these leading works the study of the self-interaction of a charge was extended to other geometries.  A notable result was the self-force on a charge in the vicinity of a straight cosmic string arising from symmetry breaking in a system composed by a complex scalar field coupled to a gauge field \cite{lin}. The associated geometry is locally flat but includes a deficit angle determined by $\mu$, the mass per unit length of the string \cite{vilgothis}. The self-force in this case points outwards from the cosmic string and is proportional to $\mu/r^2$. This non null self-force in a locally flat background is of great interest because it shows how the global properties of a manifold (in this case, the existence of a deficit angle) are revealed by the electromagnetic field of the charge. 
 
The  results described above together with the calculation of the self-force on a point charge in a wormhole space-time \cite{kb}, which turned out to be attractive, i.e. towards the wormhole throat, suggested the possibility of detecting thin-shell wormholes by means of electrostatics. Differing from well-known wormholes of the Morris--Thorne type \cite{mo} which are supported by non localized exotic matter, thin-shell wormhole geometries are supported by a shell of exotic matter located at the wormhole throat \cite{visser}. The throat connects two (equal or different) geometries which can be those of other astrophysical objects. For example, Schwarzschild thin-shell wormholes connect two  exterior (that is, beyond the horizon) non charged black hole space-times; hence the geometry at each side of the throat is locally identical to the exterior of a black hole geometry. However, the topology of the wormhole geometry is non trivial, thus the  global properties are essentially different in each case.

Inspired by the previous discussion, the philosophy of the present work  is that global aspects, such as the existence of a throat or not, may be revealed by studying electrodynamics, in particular, by the electrostatic self-force on a point charge. In our recent article \cite{eoc12}  this proposal was developed and applied it to the case of wormholes with a cylindrical throat which are mathematically constructed by removing the regions $r<a$ of two gauge cosmic string manifolds and pasting the two regions at $r\geq a$. The self-force on a charge in the cylindrical wormhole geometry was calculated, and compared it with the self-force on a charge in the vicinity of a gauge cosmic string. The result is that the force in the wormhole case can be attractive or repulsive depending of the position of the charge; this result would then allow an observer to distinguish between two geometries which are locally equal. The same argument was 
applied to the Schwarzschild case by the authors in \cite{faltante}. Related works are also \cite{khusnutdinov4}-\cite{lipatova}.

It should be mentioned that there exist some works related to these ideas. For instance, in \cite{gralla}, the authors considered a minimally coupled scalar charge and an electromagnetic charge when a Schwarzschild black hole interior is replaced by a material body and found that the leading term in a large-$r$ expansion of the force was independent of the central body type. Nevertheless, when the scalar charge is not minimally coupled, the self-force is dependent on the composition of the body. Another work in the same line is \cite{poisson}, where a spherical ball of perfect fluid in hydrostatic equilibrium with rest mass density and pressure related by some polytropic equations of state is considered. The authors found that the leading term of the force is universal and does not distinguish the internal body structure, but the next-to-leading order term is sensible to the equation of state. Thus the self-force distinguishes the body composition.

In the present work our studies about electrostatics in black hole geometries is extended to the three dimensional case, which is not a completely explored area. The natural candidates to consider are the BTZ black hole and wormhole. These geometries, although tridimensional and non realistic, have several features
 that makes them an interesting test laboratory. First of all, both have a negative cosmological constant $\Lambda<0$, which corresponds to an attraction instead of repulsion. On the other hand, their metric is not asymptotically flat, but asymptotically anti De Sitter. In addition there exist a radial coordinate $r$ such that the circles of $r$ constant have perimeter $2\pi r$, but the relative distance $d(r,r+1)$ between points located on the same radial line at positions $r$ and $r+1$ goes to zero as $r\to \infty$. This behavior is not characteristic for simple black holes in higher dimensions and is a consequence of the attractive cosmological constant term. This behavior has consequences on the boundary conditions of the electrostatic problem. These consequences will be elaborated in the paper. 
 
 It should be remarked that there are no summation formulas allowing to find a closed analytic for the self force, thus it is given in series expansion. But the series expansion of the singular part of the electrostatic field is rather complicated, and can be achieved by certain specific parameterization of the radial distance, which is explained in detail in section 7.1.  These tricks result in a series expansion for the total force that is convergent to the real one. However, we argue that the convergence to the real solution is not uniform, in other words, as larger the coordinate $r_0$ of the charge becomes, the larger the quantity of terms that has to be summed in order to approximate the self force at that point. This results in a problem when truncating the series at $r\to \infty$.
 \\

The present work is organized as follows. Section 2 contains a brief description of the BTZ black hole geometry. Section 3 contains a description of electrostatic equations in the geometry and the problems for fixing the boundary conditions for the physical solutions. Section 4 contains the expression for the electrostatic field for the BTZ black hole and wormhole. Section 5 contains a review of Synge calculus, which is a  relevant tool for calculating the singular part of the electrostatic Green function of the geometry. In section 6 this singular part is calculated for the BTZ local geometry. In section 7  a series expansion for the electrostatic self force is calculated, and the problems  mentioned above about the convergence at the asymptotic boundary is described. Section 8 contains and 9 contains interpretations of the results, which are rather non trivial. Section 10 is a summary of the obtained results. 
In the appendix there are collected some useful formulas which are applied along the text.

\section{The BTZ black hole}
Since the appearance  of the seminal works \cite{jackiw}-\cite{witten}, General Relativity in (2+1) dimensions became a widely analyzed model for exploring classical and quantum gravity, since it is recognized as a useful laboratory for studying  real system properties in (3+1) dimensions. In (2+1) dimensions GR there is no newtonian limit and there are no local degrees of freedom (that is, there are no gravitational waves in the classical theory or gravitons in the quantum theory). It came as a surprise for some then when the black hole BTZ solution was found \cite{btz}. This black hole has important differences with the Schwarzschild and Kerr black holes: it is asymptotically anti De Sitter and not asymptotically flat, and does not have any curvature singularity at the origin. Nevertheless, it is clearly a black hole: it has an even horizon and (in the rotating case) an internal horizon, and thermodynamical properties similar to black holes in (3+1) dimensions.

The BTZ solution is well known, but in order to fix the conventions  a short
description of the local and global properties of the geometry will be given. The discussion is not exhaustive, but focused in the aspects that are more important for the present work.
\subsection{Parameters of the solution}

The BTZ black hole is a solution of the Einstein field equations in  (2+1) dimensions with cosmological constant $\Lambda<0$, which bears some similarities with black hole
solutions in four dimensions \cite{btz}-\cite{btz2}. These are for instance as the presence of a event horizon, an inner horizon and an ergosphere. Also, it has a non vanishing Hawking temperature 
and interesting thermodynamical properties \cite{carlip}. Despite these similarities, there are several differences between BTZ black holes and
Schwarzschild or Kerr ones. The later are asymptotically flat, the BTZ solution instead is asymptotically anti De Sitter. Furthermore, the BTZ solution does not have a singularity at the origin.
But since the BTZ  structure is simpler than its four dimensional counterparts, it may be a good testing laboratory for making exact calculations.

The local form of the BTZ solution is well known, but in order to fix the notation  a brief review of these solutions will be given. Starting with the three dimensional action \cite{btz2}
\begin{equation}\lb{accion}
 I=\frac{1}{2\pi} \int \sqrt{-g} [R+2l^{-2}]d^2x dt + B,
\end{equation}
with $B$ is a surface term and $l$ is related to the cosmological constant by $-\Lambda=l^{-2}$, it follows that the extremal solutions corresponding to $g_{\mu\nu}(x,t)$ variations are given by the Einstein equations
\begin{equation}
 R_{\mu\nu}-\frac{1}{2}g_{\mu\nu}(R+2l^{-2})=0,
\end{equation}
which, in three dimensions only, completely determine the Riemann tensor as
\begin{equation}
 R_{\mu\nu\lambda\rho}=-l^{-2}(g_{\mu\lambda}g_{\nu\rho}-g_{\nu\lambda}g_{\mu\rho}).
\end{equation}
This solution corresponds to a symmetric space with negative curvature. If one restricts the attention to solutions possessing a rotational Killing vector $\partial / \partial \theta$ 
and a time like Killing vector $\partial / \partial t$, then by an specific choice of the radial coordinate it follows that
the line element is given by
\begin{equation}\lb{bitiz}
 ds^2=-N^2 dt^2 + N^{-2}dr^2+ r^2(N^{\theta} dt + d\theta)^2,
\end{equation}
with $N^2(r)$ and $N^\theta (r)$ the following radial functions
\be\lb{ene1}
N^2(r)=-M + \frac{r^2}{l^2} + \frac{J^2}{4r^2}, 
 \ee 
 \be \lb{ene2}
 N^\theta(r)=-\frac{J}{2r^2}.
\ee
The range of the coordinates is $-\infty <t< \infty$, $0<r<\infty$ and $0\leq \theta \leq 2\pi$. The two integration constants in
 (\ref{ene1}) and  (\ref{ene2}) are $M$ and $J$ and correspond to the mass  and angular momentum of the solutions  respectively \cite{btz2}.

The BTZ space time is not asymptotically flat. For large radial values $r \to \infty$ the metric becomes
\begin{equation}
 ds\to -(\frac{r}{l})^2 dt^2 + (\frac{r}{l})^{-2}dr^2+ r^2 d\theta ^2,
\end{equation}
which shows that this solution is asymptotically anti De Sitter.

The function $N(r)$ vanish for the following two $r$ values
\begin{equation}
 r_\pm=l \bigg[\frac{M}{2} \bigg(1\pm \sqrt{1-\bigg(\frac{J}{Ml}\bigg)^2} \bigg) \bigg]^{1/2}.
\end{equation}
The value $r_+$ corresponds to the horizon of the black hole. It exist if the following inequalities are satisfied
\begin{equation} \lb{condagujero}
 M>0, \; \; \; \; \; \; \; \; \; |J|\leq Ml.
\end{equation}
In the extreme case $|J| =Ml$, both roots of $N^2=0$ coalesce into one.
The mass $M$  and the angular momentum $J$ can be expressed in terms of $r\pm$ as
\begin{equation}
 M=\frac{r_+^2 + r_-^2}{l^2}, \; \; \; \; \; \; \; \; \;  J=\frac{2r_+r_-}{l}.
\end{equation}
For large $l$ the exterior horizon tends to infinite and only the interior remains. The vacuum state is obtained when 
the black hole disappear, and this corresponds to take the horizon radius to zero. This is equivalent of taking
 $M \to 0$, which implies $J \to 0$ due to (\ref{condagujero}).
In this case
\begin{equation}
 ds_{vac}^2=-(\frac{r}{l})^2 dt^2 + (\frac{r}{l})^{-2}dr^2+ r^2 d\theta ^2.
\end{equation}
When $M$ becomes negative, the solutions studied in \cite{deser} are found. The conical singularity that they posses is a naked
one, such as the one as a black hole with negative mass in (3+1) dimension. Such value should be excluded from the spectrum.
Nevertheless there exists an exceptional case.
When  $M=-1$ and $J=0$ the naked singularity disappears. There is no horizon in this case, but also no singularity to hide. The solution corresponding to this regime is \begin{equation}
 ds^2=- (1+ (r/l)^2)dt^2 + (1+(r/l)^2)^{-1}dr^2+ r^2 d\theta^2,
\end{equation}
and is AdS as well.

\subsection{Particular properties of the non rotating geometry}
In this section some properties of the BTZ black hole will be pointed out, which will be relevant 
when analyzing the electrostatic properties of charges in the geometry. In the present work  the non rotating case $J=0$ will be only considered.
The rotating case is leaved for a forthcoming publication.

An observation which will be of importance for interpreting the results of the present work is  that, in the non rotating BTZ geometry, the distance $d(r, r+1)$ between two points with the same $\theta$ values and lying on the circles $ r$ and $ (r+1)$  decreases
when $r$ increases. To see this, consider for simplicity the case $M=l=1$. Then the distance from a point with coordinate $r$ to the horizon $r_h$ is
\begin{equation}\lb{diston}
 d=\log(r+\sqrt{r^2-1}),
\end{equation}
which can be inverted to give $r=\cosh (d)$. When $d>>1$ it follows that $r\sim e^d$. If two points 
lying on the same line $\theta=\theta_0$ are at positions $r$ and $r+\delta r$, then the last formula gives that
\begin{equation}
 \delta r= e^{d+\delta d}-e^d= e^d(e^{\delta d}-1),
\end{equation}
which leads to
\begin{equation}\lb{future}
 \delta d=\log \bigg(1+\frac{1}{e^d}\bigg)= \log \bigg(1+\frac{1}{r}\bigg).
\end{equation}
From here it is seen that  for $r>>1$, which implies going far from the horizon $d>>1$, the true distance $\delta d$ between these points goes to zero $\delta d\rightarrow 0$. 
This particularity holds for other values of $M$ and will play a significant role in the interpretation of our results.

\section{The equations of electrostatics in BTZ space times}
In the present section  the Maxwell equations corresponding to a static charge $q$ located at $r_0$ and $\theta_0=0$ in a BTZ black hole will be derived.
The effect of the curved geometry is to deform the field lines and, as a consequence, the charge $q$ experience a self-force due to its own electric field.
As it will be  shown below, the Maxwell equations are separable in this case. Nevertheless the analysis of the physical and unphysical solutions is more 
involved than in the flat case due to the particularities of the geometry mentioned in the previous section, in particular, the behavior (\ref{future}).
 These aspects are carefully examined below and a criteria for discarding unphysical solutions is found.

\subsection{Separation of variables}
The Maxwell equations in three dimensional curved space times in natural units are given by \cite{ll} 
\be\lb{poto}
\frac{1}{\sqrt{-g}}\partial_\alpha( \sqrt{-g}g^{\mu\alpha}g^{\nu\beta}F_{\mu\nu})=2\pi j^\alpha,
\ee
$$
\epsilon_{\beta\gamma\delta}\partial_\beta F_{\gamma\delta}=0.
$$
Here $F_{\mu\nu}=\partial_{\mu}A_{\nu}-\partial_{\nu}A_{\mu}$ is the field strength tensor, $A_\mu$ is the vector potential and $j^\alpha$ the three current in the geometry.
For an static charge $q$ in front of the non rotating geometry one has that $$j^t = \frac{q}{r}\delta(r-r_0)\delta(\theta-\theta_0),$$ with $(r_0,\theta_0)$ the coordinates of the position of the charge.
The Maxwell equations (\ref{poto}) in this situation reduce to
\be
-\partial_r (rF_{tr})+\frac{r}{Mr^2-\frac{r^4}{l^2}}\partial_\theta  F_{t\theta}=2\pi r j^t,\ee
\be\lb{zri}
\partial_\theta \bigg[\frac{1}{r}(M-\frac{r^2}{l^2} )F_{r \theta}\bigg]=0,\ee
\be
\partial_r \bigg[\frac{1}{r}(M-\frac{r^2}{l^2}) F_{r \theta}\bigg]=0.\ee
Assuming that the vector $A_\mu$ is time independent it follows from these equations that there exist a gauge in
which only the component $A_t$ is non zero, and the three (\ref{zri}) reduce to the following single equation 
\be\lb{ecat}
-\partial_r (r \partial_r A_t)+\frac{r}{Mr^2-\frac{r^4}{l^2}}\partial^2_\theta  A_t=2\pi r\, j^t.
\ee
Outside the position of the charge this equation is homogeneous and can be solved by variable separation by postulating
\be
A_t (r, \theta) =R(r)\Theta(\theta).
\ee
When this is inserted into (\ref{ecat}) it is obtained that
\be
\Theta(\theta)= \exp[i n(\theta-\theta_0)],\ee
where $n$ is an integer due to the periodicity on $\theta$, and the following equation for $R(r)$ 
\be\lb{e}
r^2(1-\frac{r^2}{Ml^2}) \partial^2_r R(r) + r(1-\frac{r^2}{Ml^2})\partial_r R(r) +\frac{n^2R(r)}{M}=0.
\ee
By further defining the horizon radius $r_h ^2=Ml^2$ and making the variable change 
$r^2 \rightarrow x r_h ^2$ it is transformed into
\be
x^2 (x-1) \partial^2_x R(x) + x (x-1) \partial_x R(x)-\frac{n^2R(x)}{4M}=0. \ee
This equation has two regular singular points, which corresponds to the horizon $x=1$ and the infinite $x\rightarrow \infty$. In order to analyze the behavior at the infinite it is 
customary to make the change of variables $x \rightarrow \frac{1}{u}$
which transforms the last equation into
 \be\lb{hip}
u(1-u) \partial^2_u R + (1-u) \partial_u R-\frac{n^2R}{4M}=0. 
\ee
The equation (\ref{hip}) is a particular case of the hypergeometric one
\be\lb{hipo}
u (1-u) R''+[\gamma -(1+\alpha+\beta)u] R'-\alpha\beta R=0,
\ee
corresponding to the particular values $$\gamma=1, \qquad \alpha+\beta=0, \qquad  \beta^2=-n^2/4M.$$ It is important to remark that the
change of variables $u = \frac{1}{x}$ just performed is regular in the exterior region $r>r_h$ of the black hole, which is the region which we are interested in.

\subsection{Solutions centered around the infinite}
Having derived the equation (\ref{hipo}) which characterizes the radial behavior of the electrostatic potential $A_t$, the next task is 
to find their solutions. Since it is a linear equation of second order, it has two independent solutions. The most elementary one, which is centered around $u=0$ ($r\to \infty$), is given by  the hypergeometric series \cite{whitaker}-\cite{gradystein}
\be\lb{sol1}
f_n=\,_2F_1(\alpha_n,-\alpha_n;1;u) = \sum_{m=0}^\infty \frac{(\alpha_n)_m (-\alpha_n)_m}{(m!)^2}u^m,
\ee 
where
$$
\alpha_n= \frac{in}{2\sqrt{M}},
$$
and the Pochhammer symbols $(a)_n$ are defined by
$$
(\alpha)_m =\alpha (\alpha+1) (\alpha+2)\ldots (\alpha +m-1),\qquad (\alpha)_0=1.
$$
The elementary D' Alembert principle shows that this series is convergent for  $\arrowvert u \arrowvert<1$. Besides, when 
\begin{equation}\lb{Recab}
Re(\gamma-\alpha-\beta)>0, 
\end{equation}
the series is also convergent in $\arrowvert u \arrowvert=1$ \cite{sneddon}.  This condition is satisfied in our situation since $\beta=-\alpha$ and $\gamma=1$. 
The zone $\arrowvert u \arrowvert>2$ corresponds to the inner part of the black hole, which is of no interest to us.

The transformation $R\rightarrow u^{1-\gamma}R$ applied to  (\ref{hipo})  transforms it into another hypergeometric equation but induces a parameter transformation $(\alpha, \beta, \gamma)\rightarrow (\alpha-\gamma+1, \beta-\gamma+1, 2-\gamma)$. Therefore, in general, the function
\be
g_n=u^{1-\gamma}F(\alpha-\gamma+1, \beta-\gamma+1, 2-\gamma, u),\ee  
 is also a solution of (\ref{hipo}). Nevertheless, when $\gamma=1$, as in our case, this solution is equivalent to $f_n$, and gives no new information. In these particular cases, a new solution is obtained by postulating a
 series of the form
 \be
g_n=f_n \log u + \sum_{n=0}^\infty c_r u^r,\ee  
with $c_r$ constant coefficients to be determined. By inserting this into  (\ref{hipo}) the following recurrence for $c_r$ is obtained
\be\
(r+1)^2 c_r-r(\alpha+\beta+1)c_{r+1}+\frac{(\alpha\beta-\alpha-\beta)(\alpha)_r(\beta)_r}{r!(r+1)!}=0,\ee  
which, when solved explicitly, gives the following solution
\be\lb{solu2}
g_n=\,_2F_1(\alpha_n,-\alpha_n;1;u) \log u + \sum_{m=0}^\infty \frac{(\alpha_n)_m (-\alpha_n)_m}{(m!)^2} \, u^m \, S_{n,m},
\ee  
with
\be\lb{tar}
S_{n,m}= \sum_{k=0}^{m-1}\bigg(\frac{1}{k+\alpha_n}+\frac{1}{k-\alpha_n}-\frac{2}{k+1}\bigg).
 \ee
Alternatively, this second solution may be though as the limit \cite{sneddon}
\be
g_n=\lim_{\gamma \to 1} \frac{u^{1-\gamma}\,_2F_1(\alpha-\gamma+1, \beta-\gamma+1, 2-\gamma, u)-\,_2F_1(\alpha, \beta,\gamma, u)}{\gamma-1}.
\ee  
An important property of hypergeometric functions is the following \cite{whitaker}
\be\lb{prophiper}
\,_2F'_1(\alpha, \beta,\gamma, u)=\frac{\alpha\beta}{\gamma}\,_2F_1(\alpha+1, \beta+1,\gamma+1, u),
\ee  
which express it derivatives in terms of other hypergeometric functions. From this property and the definition (\ref{sol1})-(\ref{solu2}) for $f_\nu$ and $g_\nu$
 it follows that
 \be\lb{dy1}
     \frac{\partial f_n}{\partial r}= \frac{2\alpha_n^2 u^{3/2}}{r_h}\,_2F_1(\alpha_n+1,-\alpha_n+1,2,u),
     \ee
\be\lb{dy2}
\frac{\partial g_n}{\partial r}= -\frac{2}{r_h}\, u^{3/2} \bigg[ u^{-1} \,_2F_1(\alpha_n,-\alpha_n,1,u)-\log (u) \, \alpha_n^2 \,_2F_1(1+\alpha_n,1-\alpha_n,2,u) 
\ee
$$
  + \sum_{m=1}^\infty \frac{(\alpha_n)_m (-\alpha_n)_m}{(m!)^2} m \,  u^{m-1} S_{n,m} \bigg].
 $$
 In deriving this formulas the definition $u=r_h^2 r^{-2}$ was taken into account.  These formulas will be useful when evaluating the electrostatic field of the charge $q$ as derivatives of the potential $A_t$.
 
The behavior when $r  >> r_h$ (which corresponds to $u \to 0$)  of the solutions is directly inferred from their definition, the result is
\begin{equation}\lb{ye1}
 f_n \to 1,\qquad u\to 0,
\end{equation}
\begin{equation}\lb{ye2}
g_n\sim \log(u) \to - \infty,\qquad u\to 0.
\end{equation}
The behavior of their derivatives for large $r  >> r_h$ is
inferred by taking into account the following elementary limits
\be\lb{lim2}
\lim_{u \to 0} F(\alpha_n, -\alpha_n, 1,u)=1,
\ee
\be\lb{lim1}
     \lim_{u \to 0} \log(u) u^n=0,  \,\,\,\, \,\, \forall \; n>0,
       \ee    
    \be\lb{lim3}
   \lim_{u \to 0} F(\alpha_n+1, -\alpha_n+1, 2,u)<\infty,
       \ee
the last limit follows from the fact that any hypergeometric function is convergent at $u=0$. These limits, together with (\ref{dy1})-(\ref{dy2}) show that
\be\lb{ho1} \frac{d}{dr}f_n \sim\ \frac{1}{r^3}, \qquad r\to \infty \ee
\be\lb{ho2}\frac{d}{dr}g_n \sim \frac{1}{r},\qquad r\to \infty. \ee
Thus none of the derivatives of the solutions is divergent at the asymptotic region. Note that this behavior is in contrast with ordinary electrodynamics in $R^2$ or $R^3$, where 
there always exist a solution whose electrostatic field is divergent at infinite and is discarded in physical problems. This fact will play a crucial role in our analysis.

Consider now the behavior near the horizon $r\to r_h$ or $u \to 1$.  Both solutions (\ref{sol1}) y (\ref{solu2}) are both finite for $\arrowvert u\arrowvert=1$ since
$$f_n \to \,_2F_1(\alpha_n,-\alpha_n,1,1) = \frac{1}{\Gamma(1-\alpha_n)\Gamma(1+\alpha_n)} ,$$
\be\lb{low}g_n \to \sum_{m=0}^\infty \frac{(\alpha_n)_m (-\alpha_n)_m}{(m!)^2} S_{n,m}<\infty,\ee
the second inequality follows from the D«Alembert criteria for  series. More specifically, the function $S_{n,m}$ defined in (\ref{tar}) can be approximated by an integral
whose result is
$$
S_{n,m}\sim \log\frac{m^2-\alpha_n^2}{(m+1)^2}-\log(-\alpha_n^2),
$$
and remembering that $\alpha_n$ is purely imaginary it follows that $S_{n,m}<\infty$ for all $m$. Therefore (\ref{low}) is
\be\lb{low2}
\arrowvert g_n\arrowvert  \to \arrowvert \sum_{m=0}^\infty \frac{(\alpha_n)_m (-\alpha_n)_m}{(m!)^2} S_{n,m}\arrowvert <\arrowvert S_{n, m}^{max}\sum_{m=0}^\infty \frac{(\alpha_n)_m (-\alpha_n)_m}{(m!)^2}\arrowvert 
\ee
$$
=\arrowvert S_{n, m}^{max}  \,_2F_1(\alpha_n, -\alpha_n,1,1)\arrowvert <\infty,
$$
where in the last step (\ref{lim2}) and the definition (\ref{sol1}) has been taken into account. This shows that (\ref{low}) holds.
The derivatives $f_n$ with respect to $r$ involve functions of the form $ _2F_1(1+\alpha_n,1-\alpha_n,2,u) $, which do not satisfy (\ref{Recab}). This means that $\frac{d}{dr}f_n$ is divergent in the horizon $u=1$. The analysis for $\frac{d}{dr}g_n$  is more involved. The first term  (\ref{dy2}) is convergent. The second is also convergent, but the third is divergent. Thus the final result is that
\be\lb{ho5} \frac{d}{dr}f_n \to \infty, \qquad r\to r_h\ee
\be\lb{ho6} \frac{d}{dr}g_n \to \infty,\qquad r\to r_h.\ee
It is not easy to work with functions with this divergent behavior. Fortunately, there exist a linear combination $\alpha f_n+\beta g_n$ of both solutions whose derivative is convergent at $\arrowvert u \arrowvert=1$. This combination can be found by considering the set of solutions centered at the horizon $u=1$.

\subsection{Solutions centered around the horizon}
As it was mentioned above, the equation (\ref{hipo}) has three regular singular points. The solutions ($f_n$,  $g_n$) found in the previous section are centered around the regular singular point $u=0$, which corresponds to the asymptotic region, and are convergent in the interval $ 0 < \arrowvert u \arrowvert < 1$. Consider now a set of solutions ($h_n , k_n$)  centered
around the regular singular point $u=1$. These solutions will be convergent, as it will be shown below, in the interval $0 < u< 2 $, in particular for $ r_h < r < \infty$. This means that in the overlapping region $ 0 < u  < 1$, which is $ r_h < r < \infty$ both sets  ($f_n$,  $g_n$) and  ($h_n$,  $k_n$) constitute a basis of solutions, therefore there should exist  a relation of the form 
\be\lb{linea1}
h_n = a \, f_n + b \, g_n, 
\ee
\be\lb{linea2}
k_n = c \, f_n + d \, g_n,
\ee
valid in the overlapping region, with $a,b,c,d$ some constant coefficients. These coefficients can be found by evaluating these equalities and its first derivatives in an arbitrary point $r_0$ inside the overlapping region, the 
result is
\be\lb{coe}
a=\frac{W(h_n, g_n)(r_0)}{W(f_n, g_n)(r_0)},
\qquad
b=\frac{W(h_n, f_n)(r_0)}{W(f_n, g_n)(r_0)}.
\ee
\be\lb{coe2}
c=\frac{W(k_n, g_n)(r_0)}{W(f_n, g_n)(r_0)},
\qquad
d=\frac{W(k_n, f_n)(r_0)}{W(f_n, g_n)(r_0)}.
\ee
Here $W(f,g)=g(r)\partial_r f(r)-f(r)\partial_r g(r)$ is the wronskian of the two functions $f_n$ and $g_n$. Naturally, the value of $a,b,c,d$ does not depend on the choice of $r_0$.

A method for finding the solutions ($h_n , k_n$) is the following. Consider the change of variables $s=1-u$. The equation (\ref{hipo}) in this variable takes the form
\be\lb{hipi}
s (1-s) \tilde{R}''(s) + s \tilde{R}'(s) - \frac{n^2}{4M}\tilde{R}(s) = 0. 
\ee
Clearly, the solutions of (\ref{hipo}) around $ u = 1$ correspond to solutions of (\ref{hipi}) around $ s = 0$. The equation (\ref{hipi}) is an hypergeometric one with parameters $$\gamma = 0,\qquad \alpha_n = \frac{i n}{2\sqrt{M}},\qquad \beta_n = \frac{-in}{2\sqrt{M}}.$$ Its solutions are given by
\be
\lb{solh} h_n(r) = s \,_{_2}F_{_1}{(1+\alpha_n,1-\alpha_n,2,s)}, \ee
\be
\lb{solk} k_n(r) = s \ln(s)\,_2F_1(1+\alpha_n,1-\alpha_n,2,s)  + s \sum_{m=1}^\infty \frac{(1+\alpha_n)_m (1-\alpha_n)_m}{(2)_m (m!)}\,s^m \tilde{S}_{n,m} - \frac{1}{\alpha_n^2},
\ee
where $\alpha_n$ is the same as before and
\be\lb{al}
 s= 1 - u = 1 - \frac{r_h^ 2}{r^2}.
 \ee
 Note that non of the solutions (\ref{solh})-(\ref{solk}) are simply hypergeometric functions. This happens for some special choice of parameters, such as in our case.
 In addition 
 $$
 \tilde{S}_{n,m}= \sum_{k=0}^{m-1}\bigg(\frac{1}{k+1+\alpha_n}+\frac{1}{k+1-\alpha_n}-\frac{1}{k+1}-\frac{1}{k+2}\bigg) 
 = \psi(m+1+\alpha_n)-\psi(1+\alpha_n) 
 $$
 $$
 +\psi(m+1-\alpha_\nu)-\psi(1-\alpha_\nu)- \psi(m+2)+\psi(2)-\psi(m+1)+\psi(1),
$$
has been introduced, with $\psi(a)$the digamma function. This new variable change maps the exterior region of the black hole to $ 0~\leq~s~\leq~1 $. The line $ s = 0 $ corresponding to the event horizon and $ s = 1$ corresponds to the asymptotic region.

The derivatives with respect to $r$ are given by
\be
\lb{deriv solh} \frac{d}{dr}h_n  = \frac{2}{r_h} \left(1-s \right)^{3/2} \bigg[ \tilde{F_n}(s) + s \tilde{F_n}'(s) \bigg] \ee
\be
\label{deriv solk} \frac{d}{dr}k_n  = \frac{2}{r_h} \left(1-s \right)^{3/2}  \bigg[ \left(1 + \ln s \right) \tilde{F_n}(s) + s \ln s \tilde{F_n}'(s) \ee
$$
+ \sum_{m=0}^\infty \frac{(1+\alpha_n)_m (1-\alpha_n)_m}{(2)_m m!} (m+1) \,  s^m S_{n,m} \bigg] ,
$$
where the following notation has been introduced for simplicity \be\lb{no1}
\tilde{F_n}(s) \equiv \;_2F_1(1+\alpha_n,1-\alpha_n,2,s) ,\ee
\be\lb{no2}
\tilde{F_n}'(s)  \equiv \frac{(1-\alpha_n^2)}{2} \; _2F_1(2+\alpha_n,2-\alpha_n,3,s).
\ee
The behavior far from the horizon $ r >> r_h $ or $ s \to 1$ is not easily seen from (\ref{solh})-(\ref{solk}) or  (\ref{deriv solh})-(\ref{deriv solk}). For instance, factor $(1-s)^{3/2}$  in (\ref{deriv solh})- goes to zero at the infinite but the combination inside the parenthesis is divergent, so there is  a $0.\infty$ ambiguity.  But the result of this limit can be inferred from  (\ref{linea1})-(\ref{linea2}), since $h_n$ and $k_n$ are linear combinations of the functions $f_n$ or $g_n$, which are centered at the infinite. Since $g_n\sim -\log(r)$ is divergent at the infinite it follows that 
\be\lb{lis} h_{n} \sim  \log(r)\rightarrow \infty,\qquad r\to \infty,\ee
\be\lb{lis2} k_{n}\sim  \log(r)\rightarrow\infty,\qquad r\to \infty.\ee
The behavior from the derivatives follows from (\ref{ho1})-(\ref{ho2}), since  $k_n$ and $h_n$ are linear combinations of $f_n$ and $g_n$ and the last two have finite derivatives at the asymptotic region. From this simple fact it follows that
\be\lb{ho3}
\frac{d}{dr}h_n\sim\frac{1}{r}  \rightarrow 0,\qquad r\to \infty,
\ee
\be\lb{ho4}
\frac{d}{dr}k_n\sim\frac{1}{r}  \rightarrow 0\qquad r\to\infty.
\ee
On the other hand, the behavior of the new solutions (\ref{solh}) and (\ref{solk}) at the horizon  $s \to 0$ is directly seen from its definitions, it is given by
\be\lb{beha1}h_{n} \sim s F(s) \to  0,\qquad s\to 0 \ee
\be\lb{beha2} k_{n} \to s \log(s)F(s)- \frac{1}{\alpha_n^2}\to  -\frac{1}{\alpha_n^2} ,\qquad s\to 0.\ee
The behavior of their derivatives is
\be\lb{beha3}
\frac{d}{dr}h_n \sim \frac{2}{r_h}(F(s)+s F'(s))\to  \frac{2}{r_h} ,\qquad s\to 0
\ee
\be\lb{beha4}
\frac{d}{dr}k_n\sim  \frac{2}{r_h}[(1+\log s)F(s)+s \log s F'(s)+s] \to -\infty,\qquad s\to 0.
\ee
Therefore we have reached our goal namely, to find a basis for which one of the eigenfunctions behave regularly at the horizon. 
It will be more convenient for our purposes to work with solutions satisfying  (\ref{beha1})-(\ref{ho4}) than using the ones satisfying (\ref{ye1})-(\ref{ho6}). For this reason the following calculations will be referred to the set
constituted by $h_n$ and $k_n$.

\subsection{The unphysical solutions}
After elucidating the behavior of the solutions of the equation (\ref{ecat}) for the potential $A_t$, the next step is to discuss the boundary conditions for the electrostatic problem. The particularities of the BTZ geometry discussed in previous sections make the analysis different than in ordinary electrostatic in flat spaces, since the geometry is not asymptotically flat. In addition the behavior of the distance $d(r,r+1)$ given in (\ref{future}) does not hold in a flat geometry. As a consequence of  this behavior, when the usual boundary conditions of electrostatic in flat space are applied to this case, the electric field is not uniquely defined. This is an artifact which suggest that the new types of boundary conditions should be considered for the non asymptotically flat geometry.

The problems described above can be illustrated with an heuristic argument as follows. Consider a perfect dipole in flat space $R^3$, constituted by two charges $q$ and $-q$ separated by a distance $d$. An elementary result in electrostatic states that the dipolar momenta $p$ of such configuration is independent on the origin $O$ of the coordinates.  Thus this dipolar moment is the same near the origin or far away from it. This situation is radically different in the BTZ geometry. As it was discussed in (\ref{future}), the distance $d(r)$ between two points lying and on circle of radius $r$ and $r+1$ and on the same line $\theta=\theta_0$  tends to zero when $r\to\infty$. Consider now two charges $q$ and $-q$ located at these points. If  these charges are translated simultaneously in radial direction to $r\to\infty$, since their mutual distance $d(r)\to 0$, these charges become superposed one onto the other. It may then seem plausible that all the multipolar momenta tend to zero in this limit. The same reasoning holds for radially directed finite charged lines with total charge equal to zero.

The discussion given above suggest that one can send to the asymptotic region  any finite number of radially directed neutral configurations, which will seem to disappear at the infinite. But if an infinite number of configurations is sent, then the result is ambiguous, since the resulting multipoles are an indetermination of the form $0.\infty$. 

These facts can be visualized by considering the multipole expansion of these radially directed neutral configurations. This expansion is expected to be of the form
\be\lb{shopo}
A(x,x')=\sum_{j=1}^{\infty}\frac{M^j_{i_1...i_j}(x)\sigma^{i_1}(x,x')... \sigma^{i_j}(x,x')}{\sigma^{n_j}(x,x')},
\ee
where $\sigma^i(x,x')$ is a function with reduce to the usual difference $\sigma^i=(x-x')^i$ in a flat space. In addition $M_{i_1...i_j}(x)$ are by definition multipolar momenta, and $n_j$ are positive numbers, whose explicit value is of no importance in this discussion. This is the essence of the Synge calculus \cite{synge}, to be discussed in detail in the next sections.  Here $x$ is some characteristic point of the charged body and $x'$ the observation point. The sum starts at $j=1$ since the zero multipole, which is the total charge, is assumed to be zero. 

Now, if the expansion (\ref{shopo}) is applied to the BTZ case, one encounters the following ambiguity. When a neutral radially directed configuration whose center is at $r_0$ is sent to the asymptotic region, then $r_0$ take large values and the denominator tends to zero for in an small neighborhood. This follows from the fact that $\sigma^a(r_0, r)\to 0$ when $r\to \infty$, $r_0\to \infty$ such that $r-r_0<\infty$. On the other hand, when $r_0\to\infty$ the size of the system tends to zero, this follows from the behavior of $d(r)$ in a BTZ geometry. It is plausible then that the multipoles $M^j_{i_1...i_j}$ are also zero in this limit, but this affirmation is to be taken with care. For instance, one can consider a dipole composed by two charges $q$ and $-q$, which is sent  to the asymptotic region while adding opposite charges at increasing positions, in such a way that when the dipole is centered at $r$ the charges are $q(r)$ and $-q(r)$, with $q(r)$ an arbitrary function of $r$. This function may be fixed to give a non zero value for the multipoles $M^j_{i_1...i_j}$ at infinite. In any case, if the multipoles are zero then there is an indetermination of the form $0.\infty$ for the potential (\ref{shopo}). If instead the multipoles are finite or even infinite, then  a divergence located at the infinite appears, which may give as a result a finite remanent electric field at finite $r$ values. These arguments are of course heuristic, but suggest that the appropriate boundary conditions are not that straightforward as in the flat case.

The problems discussed in the previous paragraphs are reflected in the calculations as follows. The separation of variables for $A_t$ shows that the general electrostatic potential outside the source in a non rotating BTZ black hole admits an expansion of the form
\be\lb{exp}
A_t(r,\theta)=\sum_{n=1}^\infty \exp(in \theta) (A_n h_n(r)+B_n k_n(r))+\alpha \log(r)+\beta.
\ee
In general grounds one expect the electric field $E_i$ to vanish asymptotically and to be finite at the horizon. This field is obtained by taking derivatives of $A_t$. More precisely, one expect the invariant
\be\lb{ivnar}
F_{\mu\nu}F^{\mu\nu}=-(\partial_rA_t)^2-\frac{1}{r^2 (\frac{r^2}{l^2}-M^2)}(\partial_\theta A_t)^2,
\ee
to vanish at the infinite and to be finite at the horizon. Consider the simplest configuration first, namely, the one without charges. In this case $B_n$ should be zero since the derivatives of $k_n(r)$ are infinite at the horizon by (\ref{beha4}). Thus, if $B_n$ were not zero, then the first term in (\ref{ivnar}) would be divergent. On the other hand the derivatives of $h_n(r)$ are well behaved at both the horizon and the asymptotic region. Nevertheless, its value is divergent at the infinite and since the derivative $\partial_\theta A_t$ contains terms proportional to $h_n$ it follows that  $\partial_\theta A_t\to\infty$ at the asymptotic region. But this derivative is divided in (\ref{ivnar}) by a factor which diverges when $r\to \infty$  faster than $h'_n(r)$. In fact from (\ref{lis}) it follows that
\be\lb{dvd}
\frac{1}{r^2 (\frac{r^2}{l^2}-M)}(\partial_\theta A_t)^2\sim \frac{1}{r^4}\log^2(r)\to 0,\qquad r\to \infty.
\ee
So the invariant tends to zero at the infinite and so the electric field.  The other term to be careful with is the denominator in the second term in (\ref{ivnar}), which gives a potentially
divergence at the horizon. But taking into account (\ref{beha1}) it follows that
\be\lb{dvd}
\frac{1}{r^2 (\frac{r^2}{l^2}-M)}(\partial_\theta A_t)^2\sim \bigg(\frac{1}{r^4 s}\bigg) s^2 F^2(s)\to  0,\qquad r\to r_h.
\ee
Thus, the presence of $h_n$ is not dangerous at the horizon either. The Gauss law fixes $\alpha=0$. Therefore it is concluded that in absence of charges the most general potential is
\be\lb{expl2}
A_t(r,\theta)=\sum_{n=1}^\infty A_n  \exp(in \theta) h_n(r),
\ee
with $A_n$ a arbitrary coefficients. 

At first sight, this result may lead to the awkward conclusion that there exist an electric field, corresponding to (\ref{expl2}), in absence of charge. The interpretation to be adopted in this work is that this conclusion is not true but instead, the solution (\ref{expl2}) is \emph{unphysical}, and corresponds to the electrostatic potential of "configurations in the infinite" of the type mentioned above. These configurations are characteristic in a BTZ geometry due to the pathological behavior of the radial distance $d(r, r+1)$ explained in (\ref{future}).  Therefore, in an electrostatic problem in BTZ geometry our criteria for discarding solutions will not be the request that the radial solutions goes to zero at infinite, which is customary in ordinary electrodynamics in flat space. The task of determining appropriate boundary conditions is described in  the next section.

\section{Electrostatic field of BTZ black hole and wormhole}

Having derived the eigenfunctions of the electrostatic problem in the BTZ geometry, the next step is  the calculation of the electrostatic potential 
of a point charge $q$ in front of the BTZ black hole and wormhole. 
The static charge is located at the position $r_0>r_h$, $\theta=0$. Its electrostatic potential be expressed as 
\be\lb{exp1}
A^I_t(r,\theta)=\sum_{n=1}^\infty \exp(in \theta) (A_n h_n(r)+B_n k_n(r))+\alpha\log(r)+\beta,
\ee
\be\lb{exp2}
A^{II}_t(r,\theta)=\sum_{n=1}^\infty \exp(in \theta) (C_n h_n(r)+D_n k_n(r))+\gamma  \log(r)+\delta.
\ee
The potential $A^I_t$ is the one in the region between the charge position $r_0$ and the horizon $r_h$, and the $A^{II}_t$ corresponds to the region between $r_0$ and the asymptotic boundary.  In the first region 
$B_\mu=0$ should be imposed, this is due to the fact that the derivatives of $k_\mu(r)$ and the invariant (\ref{ivnar}) would not be bounded at the horizon. For the second region, the discussion below (\ref{expl2}) suggest that if the coefficients $C_n$ multiplying $h_n(r)$ are non vanishing, then non trivial charge configurations at the infinite are turned on.  This lead us to the following.
\\

\emph{First type of boundary conditions:} In order to avoid the residual electric field above one may impose that $C_n=0$ in the region between the charge and the asymptotic boundary. Note that if this boundary conditions is imposed, then (\ref{expl2}) is automatically zero. This is expected by intuition namely, it implies the absence of electric  field in absence of charges. The resulting potential for the charge has now the form 
\be\lb{es1}
A^{I}_t(r,\theta)=\sum_{n=1}^\infty \exp(i n \theta) A_n h_n(r)+\alpha\log(r)+\beta,
\ee
\be\lb{es2}
A^{II}_t(r,\theta)=\sum_{n=1}^\infty \exp(in \theta) D_n k_n(r)+\gamma  \log(r)+\delta.
\ee
However, there is some unpleasant detail concerning this choice of boundary conditions.  First, it is not the only type of conditions that insures vanishing electric field in absence of charges. One may add to (\ref{es1})-(\ref{es2}) a solution of the homogeneous Maxwell equation with the form (\ref{expl2}), and with the coefficients $A_n$ proportional to the charge $q$, which will vanish when $q\to 0$. This shows that there is an ambiguity for the choice of the boundary conditions. 
\\

\emph{Second type of boundary conditions:} There exist an unique privileged type of boundary condition, which is seen as follows. The derivative of the functions $k_n(r)$, as shown in (\ref{ho4}), decays as $1/r$ at the asymptotic region. Instead, the derivatives of $f_n(r)$, as seen by (\ref{ho1}), decay  as $1/r^3$. In view of this, it may look strange that the asymptotic behavior of (\ref{es2}) goes like $1/r$ and not like $1/r^3$. In other words, it may be reasonable to expect that, for a localized system of charges, the potential decays as fast as possible at the asymptotic region. This is of course what happens in an ordinary problem in electrostatics. And the boundary conditions imposing $C_n=0$ and leaving to the solution (\ref{es1})-(\ref{es2}) do not respect this behavior. Therefore, one may consider the possibility of working with the base $(h_n, f_n)$ instead of $(h_n, k_n)$. In these terms the electrostatic potential in absence of charges is 
\be\lb{expera2}
A^{II}_t(r,\theta)=\sum_{n=1}^\infty \exp(in \theta) (\alpha_n h_n(r)+\beta_n f_n(r))+\gamma  \log(r)+\delta.
\ee
The requirement fast decay at the asymptotic implies $\alpha_n=\alpha=0$, otherwise the derivative of the potential would decay as $1/r$ instead as $1/r^3$. The regularity at the horizon requires $\beta_n=0$ since the derivatives of $f_n(r)$ are not bounded at the horizon. Thus these boundary conditions gives a constant potential  $\delta$ and no electric field in absence of charges as well. 
\\

It should be remarked there are many  boundary conditions giving no electric field when no charges are present, but only the second type one gives a decay of the form $1/r^3$. This follows from the fact that the radial function $f_n(r)$
is the unique between the four $f_n(r)$, $g_n(r)$, $h_n(r)$ and $k_n(r)$ which this fast decay and therefore, the addition of other of these functions in the region II will spoil this behavior.
In any case, it may be instructive to consider both types of conditions separately, and this will be done in the following subsection.

\subsubsection{The black hole electrostatics for the first type of boundary conditions}

Consider the conditions requiring $C_n=0$ first. We call this "wrong" conditions since they are non unique, although this name will be justified better when studying the charge self-force. The matching conditions for the coefficients $A_n$ and $D_n$ are the request of continuity of the potential and the request of discontinuity of the electric field when crossing the surface $r_0$ along the radial line where the charge is located. These conditions are translated into the following linear equations for
the unknown coefficients
$$
A_n h_n(r_0)=D_n k_n(r_0),
$$
$$
A_n h'_n(r_0)-D_n k'_n(r_0)=\frac{q}{r_0},
$$
$$
\gamma=-q,\qquad \alpha=0,\qquad \beta=-q\log(r_0)+\delta.
$$
The constant $\delta$ can be fixed to zero without losing generality. The solution of this system is
$$
A_n=\frac{q k_n(r_0)}{r_0W_n(r_0)},\qquad
D_n=\frac{q h_n(r_0)}{r_0W_n(r_0)},
$$
with 
\be\lb{rosko}
W(r_0)=k_n(r_0) h'_n(r_0)-k'_n(r_0) h_n(r_0),
\ee
the Wronskian of the two solutions $k_n$ and $h_n$ at the charge radial position $r_0$. Therefore the electrostatic potential is given by
\be\lb{exp01}
A^I_t(r,\theta)=-q\log(r_0)+\sum_{n=0}^\infty  \frac{q k_n(r_0)h_n(r)}{r_0W_n(r_0)} \exp(in \theta),
\ee
\be\lb{exp02}
A^{II}_t(r,\theta)=-q\log(r)+\sum_{n=0}^\infty  \frac{q h_n(r_0)k_n(r)}{r_0W_n(r_0)}\exp(in \theta).
\ee
The Wronskian (\ref{rosko}) can be calculated as follows. Consider two arbitrary linearly independent solutions $y^1_n$ and $y^2_n$ of the hypergeometric equation
$$
u(1-u) y''^i_n (u) + (1-u) y'^i_n (u) -\frac{n^2 y^i_n (u)}{4M}=0. 
$$
By multiplying the equation for $y^1_n$ by $y^2_n$ and by doing the opposite procedure for the equation for $y^2_n$, then after subtracting the resulting terms it is obtained
the following equation 
$$
u \; \widetilde W'(u) + \widetilde W(u)  = 0,
$$
for the wronskian 
$$\widetilde W(u) = y^1_n y'^2_n-y'^1_n y^2_n.$$
If the wronskian $ \widetilde W(v)$ at a point $v$ is known, then the solution of the last equation is
$$
\widetilde W(u) = \frac{ v \widetilde W(v)}{u}.
$$
The Wronskian just considered is referred to derivatives in $u$. The wronskian $W(r)$ referred to derivatives of 
$r$ is obtained by multiplying $\widetilde W$ by $u'(r)$, the result is
\be
\label{solW} W(r) = \frac{r' W(r')}{r},
\ee
with $r'$ an arbitrary fixed point. Therefore, once the wronksian at a given point $r'$ is known, its values at a generic point $r$ are determined by the last formula.
For the case in consideration, it is convenient to calculate $W(h_n, k_n)$ at $r_h$, which corresponds to $s=0$. The value follows directly from (\ref{beha1})-(\ref{beha4}), the result is
\be\lb{wrosko2}
W_n(r_h)=\frac{2 F_n(0)}{r_h\alpha_n^2},
\ee
and taking into account the definition of the hypergeometric function
$$
\,_2F_1(\alpha, \beta;\gamma;s) = \sum_{n=0}^\infty \frac{(\alpha)_n (\beta)_n}{(n)(\gamma)_n}s^n,
$$
it is concluded that
$$
\,_2F_1(\alpha, \beta;\gamma;0) =1.
$$
By this and (\ref{no1}) the wronskian (\ref{wrosko2}) takes the following form
\be\lb{wrosko}
W_n(r_h)=\frac{8M}{ r_h n^2}.
\ee
In these terms (\ref{exp01})-(\ref{exp02}) become
\be\lb{exp001}
A^I_t(r,\theta)=-q\log(r_0)+\sum_{n=1}^\infty  \frac{qn^2 k_n(r_0)h_n(r)}{8M} \exp(in \theta),
\ee
\be\lb{exp002}
A^{II}_t(r,\theta)=-q\log(r)+\sum_{n=1}^\infty  \frac{q n^2 h_n(r_0)k_n(r)}{8M}\exp(in \theta).
\ee
This is the electrostatic potential corresponding for the first type of boundary conditions.
\subsubsection{The black hole electrostatics for the second  type of boundary conditions}
The second boundary conditions, which will be called of "right" type state that the region II should be described in terms of the fast decaying radial functions $f_n(r)$, while the region I should be described by $h_n(r)$, which are regular at the horizon.
The electrostatic potential satisfying this requirement is generically
\be\lb{ep1}
A^I_t(r,\theta)=\sum_{n=1}^\infty \exp(i n \theta) A_n h_n(r)+\alpha\log(r)+\beta,
\ee
\be\lb{ep2}
A^{II}_t(r,\theta)=\sum_{n=1}^\infty \exp(i n \theta) D_n f_n(r)+\gamma  \log(r)+\delta.
\ee
The analysis of the boundary conditions at the charge position $r_0$ is completely analogous to the one performed in the previous section.  For the case $M=1$
the Wronskian constructed in terms of $(h_n(r), f_n(r))$ is given by 
$$
r_0 W_n(r_0)=\frac{2}{ \Gamma(1+\frac{in}{2}) \Gamma(1-\frac{in}{2})}.
$$
The electrostatic potential in this case is 
\be\lb{pu001}
A^I_t(r,\theta)=-q\log(r_0)+ \frac{q}{2}\sum_{n=1}^\infty \Gamma(1+\frac{in}{2}) \Gamma(1-\frac{in}{2}) f_n(r_0)h_n(r) \exp(in \theta),
\ee
\be\lb{pu002}
A^{II}_t(r,\theta)=-q\log(r)+ \frac{q}{2}\sum_{n=1}^\infty  \Gamma(1+\frac{in}{2}) \Gamma(1-\frac{in}{2}) h_n(r_0)f_n(r)\exp(in \theta).
\ee
This is the unique potential with the right discontinuity at the charge position, the right behavior at the horizon and  and with the fastest decaying conditions.

\subsection{The wormhole electrostatic field}

In order to find the electrostatic potential of a charge in front of a BTZ wormhole it is convenient to divide the space time in the following three regions 
$$
\textup{Region I:}\qquad r_- < r_g,
$$
$$
\textup{Region II: }\qquad r_g < r_+ < r_0,
$$
$$
\textup{Region III: }\qquad r_+ < r_0.\\
$$
Here $r_g$ indicates the throat position. The electrostatic solution in any of these regions is of the form 
$$
A_n^{III}= -q\log(r)+\sum_{n=1}^\infty (A_n h_n(r) + B_n k_n(r))\exp[n(\theta-\theta_0)],
$$
\be\lb{radial}
A_n^{II}= -q\log(r_0)+\sum_{n=1}^\infty (C_n h_n(r) + D_n k_n(r)) \exp[n(\theta-\theta_0)],
\ee       
$$
A_n^I= -q\log(r_0)+\sum_{n=1}^\infty (E_n h_n(r) + F_n k_n(r)) \exp[n(\theta-\theta_0)].   
$$
The coefficients $A_n,.., F_n$ are given by the boundary conditions of the problem, which are the following.
 \begin{enumerate}
        \item The potential is continuous in $r_+=r_0$,
	    \be\lb{cc1}
	    A^{III}(r_+ \rightarrow r_0^+)= A^{II}(r_+ \rightarrow r_0^-)\ee
        \item The potential is continuous in $r_-=r_+=r_g$
         \be\lb{cc2}
	    A^{II}(r_{+} \rightarrow r_g^+)= A^{I}(r_{-} \rightarrow r_g^-)\ee
        \item Continuity of the field in $r_-=r_+=r_g$
	   \be\lb{cc3}
	   \partial_{r_+} A^{II}(r_+ \rightarrow r_g^+)= -\partial_{r_-} A^{I}(r_-\rightarrow r_g^-)\ee
        \item Discontinuity of the electric field in $r_+=r_0$
           \be\lb{cc4}
	   \partial_{r_+} A^{III}(r_+\rightarrow r_0 ^+)-\partial_{r_+} A^{II}(r_+\rightarrow r_0^-)=-\frac{2\pi q}{r_0} \delta (\theta-\theta_0)\ee
        \item \lb{cc5} The correct asymptotic behavior in $\mathcal{M}_-$ and $\mathcal{M}_+$.
          \end{enumerate}
The last condition is related to the two boundary conditions discussed in the previous  section for the black hole case. For the first type of boundary conditions
\begin{equation}
 B_n=F_n=0.
\end{equation}
The other four conditions imply that
$$
A_n h_n(r_0) =C_n h_n(r_0) + D_n k_n(r_0),
$$
$$
A_n h'_n(r_0) =C_n h'_n(r_0) + D_n k'_n(r_0)+\frac{q}{r_0},
$$
$$
E_n h_n(r_g) =C_n h_n(r_g) + D_n k_n(r_g),
$$
$$
-E_n h'_n(r_g) =C_n h'_n(r_g) + D_n k'_n(r_g).
$$
This is system of four equations with four undetermined, whose solution gives
\be\lb{solw33}
A_t^{III}= -q\log(r)+\sum_{n=1}^\infty \bigg[\frac{q k_n(r_0)}{r_0W_n(r_0)} -\frac{q h_n(r_0)}{2r_0 h_n(r_g)h'_n(r_g)}\bigg[\frac{W_n(r_g)+2h_n(r_g)k'_n(r_g)}
  {W_n(r_0)}\bigg]
\bigg] h_n(r)\exp[n(\theta-\theta_0)].
\ee
Attention will be paid only for the potential in the third region, since is the one to be used for calculating the charge self-force. It can be decomposed further as
\be\lb{solw333}
A_t^{III}= A_t^{bh}-\frac{1}{2}\sum_{n=1}^\infty \bigg[\frac{q h_n(r_0)}{r_g h_n(r_g)h'_n(r_g)}+\frac{q n^2 h_n(r_0)k'_n(r_g)}{8 M h'_n(r_g)}\bigg]
 h_n(r)\exp[n(\theta-\theta_0)],
\ee
with $A_t^{bh}$ the potential corresponding to the black hole solution (\ref{exp001})-(\ref{exp002}). 

For the second type of boundary conditions the resulting potential is
\be\lb{solp333}
A_t^{III}= A_t^{bh}-\frac{1}{2}\sum_{n=1}^\infty \bigg[\frac{q h_n(r_0)}{r_g h_n(r_g)h'_n(r_g)}+\Gamma(1+\frac{in}{2}) \Gamma(1-\frac{in}{2})\frac{q h_n(r_0)k'_n(r_g)}{2 h'_n(r_g)}\bigg]
 h_n(r)\exp[n(\theta-\theta_0)],
\ee
where now $A_t^{bh}$ is the potential corresponding to the black hole solution (\ref{pu001})-(\ref{pu002}). The remaining sum is due to the effect of the throat at $r_g$, which deform the electric field lines. This shows that both geometries, which are locally the same, can be distinguished by electrostatic effects.

\section{Coincident points limits and Taylor like expansions in curved space times} 
The electrostatic potential $A_t$ for the static charge $q$ in any geometry is singular at the position where the charge is located. In a flat space, this charge does not experience
any self-force, this is clear due to the rotational symmetry of the electrostatic field. In a curved space, this argument is not true, since the non trivial curvature of the geometry deforms the electric lines and gives a net force on the charge. A seminal work about electrostatic in curved space is the one of Haddamard \cite{haddamard}, who started a program for calculating the singular part for $A_t$ in static geometries. 

The electrostatic vector potential $A_t(x,x')$ is an example of a bivector, since it depends on two arguments, the position of the charge $x$ and the position
of the observer $x'$. The self force on  the charge is determined by the behavior of $A_t(x,x')$ in an infinitesimal neighborhood of $x$, and the analogous of a Taylor expansion in a curved
space plays an important role in determining this behavior. In the present section  the main properties of these expansions are described, which requires Synge calculus \cite{synge}. The references \cite{jacobson}- \cite{five} are more detailed and contains 
more information. Nevertheless  a concise but self-contained description of the Synge calculus is given  in the following subsections.

\subsection{The Synge world function and its main properties}
Our task is to calculate the self force of a static charge in front of a BTZ black hole. This requires to calculate its electrostatic field $E$ and to substract the part that it is divergent at the position of the charge. 
There are several methods to extract this singular part. The one to be implemented here, which is better adapted to static geometries, has as a basic ingredients
the parallel propagator bitensor $g_\alpha^\beta(x,x')$ and the Synge world function $\sigma(x,x')$ \cite{synge}. To define them, consider a space time ($g$, $M$) and choose an einbein basis $e^a$ for the metric $g$ such that
$$
g_{\mu\nu}(x)=\eta_{ab}e^a_\mu(x) e^b_\nu(x).
$$
There is an $SO(n-1,1)$ freedom for choosing this basis, since $SO(n-1,1)$ rotations $R^a_b$ induce new one forms
$$
e'^a(x)=R^a_b e^b(x)
$$
which are still an einbein for the metric $g_{\mu\nu}$. In particular, since always $\nabla g_{ab}=0$ one may choose an einbein $e^a(x)$ at $x$ such that for any $x'$ lying in the injectivity radius of $x$ it is parallel transported along the unique geodesic $\gamma$ joining the two points.

Now given a vector field $A_\mu(x)$ defined at $TM_x$ one can express it in the basis $e^a(x)$ as
$$
A_\mu(x)=A_a e^a_\mu(x).
$$
If this vector is parallel transported to $x'$ along $\gamma$ then its components at that point are
$$
A_\mu(x')=A_a e^a_\mu(x'),
$$
and it follows that
$$
A_\mu(x')=g^\nu_\mu(x,x')A_\nu(x),
$$
with
\be\lb{pt}
g_{\mu}^{\nu}(x,x')=e^a_\mu(x)e_a^\nu(x').
\ee
The object (\ref{pt}) then relates the components of the vector field $A_\mu$ at $x$ and $x'$. This object is by definition
the parallel transport bitensor of the geometry.

Let us turn the attention to the Synge world function $\sigma(x,x')$. This function is defined as half of the square of the geodesic distance $d(x,x')$ between $x$
and $x'$
\be\lb{sz}
\sigma(x,x')=\frac{1}{2}d(x,x')^2.
\ee
This distance $d(x,x')$ can be represented in integral form as
\be\lb{disac}
d(x,x')=\int_0^1 \sqrt{g_{ab}\dot{x}^a\dot{x}^b}d\lambda,\qquad x(0)=x,\qquad x(1)=x',
\ee
with $\dot{x}^a$ satisfying the geodesic equation
\be\lb{gio}
\dot{x}^a\nabla_a\dot{x}^b=0.
\ee
The bivector 
\be\lb{vivo}
n_a=\nabla_a d(x,x'),\qquad n_{a'}=\nabla_{a'} d(x,x'),
\ee
constructed by taking derivatives of the distance $d(x,x')$ with respect to the initial or final point $x$ or $x'$, has unit length. This can be seen explicitly by calculating the variation of the distance 
$$\delta d=d(x+\delta x, x')-d(x,x'),$$ with $\delta x$
subject to the boundary conditions
$$
\delta x(0)=\delta x_0,\qquad \delta x(1)=0.
$$
Now the integral (\ref{disac}) represents the distance $d(x,x')$ as an action with lagrangian \be\lb{logro}\mathcal{L}=\sqrt{g_{ab}\dot{x}^a\dot{x}^b},\ee and the standard theory of Hamilton-Jacobi implies that
the last variation is
\be\lb{lava}
\delta d=n_a \delta x^a,
\ee
with $n_a$ the momentum corresponding to the coordinate $x_a$ calculated with the lagrangian (\ref{logro}), which is given by
$$
n_a=-\frac{\partial \mathcal{L}}{\partial \dot{x}_a}=\frac{ g_{ab}(x) \dot{x}^b}{\sqrt{g_{ab}\dot{x}^a\dot{x}^b}}.
$$
Clearly, this is the bivector (\ref{vivo})  and it follows from the last expression that
\be\lb{uv}
g^{ab}n_a n_b=1,
\ee
which proves that $n_a$ has unit length, as stated. In addition
\be\lb{ina}
n_{a'}(x,x')=-g_{a'}^a(x,x') n_a(x,x'),
\ee
which follows from the definition of the parallel transport bitensor. Furthermore, the norm $g_{ab}\dot{x}^a\dot{x}^b$ is constant  along a geodesic $\gamma$, this follows from the Levi-Civita condition $\nabla_c g_{ab}=0$
together with the geodesic equation (\ref{gio}). This  implies that the Synge function can be expressed in integral form as
\be\lb{sz2}
\sigma(x,x')=\frac{1}{2}\int_0^1 g_{ab}\dot{x}^a\dot{x}^b d\lambda,\qquad x(0)=x,\qquad x(1)=x',
\ee
and their derivatives $\sigma_a=\nabla_a \sigma$ also satisfy some useful identities analogous to (\ref{uv}).
One of them is
 \be\lb{bm}
 g^{ab}\sigma_{a}\sigma_{b}= g^{a'b'}\sigma_{a'}\sigma_{b'}=\sigma,
 \ee
 its proof follows directly from the action representation (\ref{sz2}) and the Hamilton-Jacobi theory.
Note that for the flat metric  $\sigma=\eta^{ab}(x-x')_a(x-x')_b$ and the identity (\ref{bm}) is immediate.
In addition the following relation takes place
\be\lb{ina2}
g^{a'}_{a}(x,x') \sigma^{a}(x,x') = -\sigma^{a'}(x,x'),
 \ee
 which is the analogous of (\ref{ina}). In the following the notation
 $\sigma_{i_1..i_n}=\nabla_1...\nabla_n \sigma$ will be employed. With this notation the formula (\ref{bm}) can be differentiated with respect to the coordinate $x$ giving that
\be\lb{trueee}
\sigma^{a'}_{a} \sigma^{a} = \sigma^{a'}.
\ee
At this point it is convenient to introduce more formally the definition of a bitensor, since this is a notion to be used recurrently in the following. 

\subsection{Taylor expansions of bitensors}
Consider an arbitrary manifold $M$ and choose two of its points  $x$ y $x'$. A bitensor 
$T_{\alpha_1..\alpha_m\alpha'_1...\alpha'_n}^{\beta_1..\beta_k\beta'_1..\beta'_l }$ is a linear application of the form
$$
T: TM_x\times...\times TM_x\times TM_{x'}\times...\times TM_{x'}\times TM^{\ast}_x\times...\times TM^{\ast}_x\times TM^{\ast}_{x'}\times...\times TM^{\ast}_{x'}\to C
$$
with $TM_{p}$ the tangent space at the point $p$ and $TM^{\ast}_{p}$ its dual. In particular  $n=l=0$ corresponds to a tensor $m$ times covariant and $k$ times contravariant.
A bitensor field is a rule that assigns to the pair of points $(x,x')\in M\times M$ the bitensor $T_{\alpha_1..\alpha_m\alpha'_1...\alpha'_n}^{\beta_1..\beta_k\beta'_1..\beta'_l }(x,x')$.
Such objets can be differentiated with respect to $x$ or $x'$. The derivative $\nabla_{a}T_{\alpha_1..\alpha_m\alpha'_1...\alpha'_n}^{\beta_1..\beta_k\beta'_1..\beta'_l }$ it is obtained by considering $x'$ frozen and taking its covariant derivative by pretending that it is an $m$ times covariant and $k$ times contravariant tensor.
As a simple example consider the object $\sigma(x, x')$, which is a  biscalar. The covariant derivative 
$\sigma_{a}(x, x')$ is a bivector and by taking successive derivatives one can construct a bitensor of arbitrary order. 

 When the points of a generic bitensor of rank two  $T_{\alpha'\beta'}(x,x')$ are close enough, one may make a Taylor like expansion of the form
\be\lb{taylor}
T_{\alpha'\beta'}(x,x')=A_{\alpha'\beta'}(x')+A_{\alpha'\beta'\gamma'}(x')\sigma^{\gamma'}(x,x')+
A_{\alpha'\beta'\gamma'\delta'}(x')\sigma^{\gamma'}(x,x')\sigma^{\delta'}(x,x')+O(\epsilon^3),
\ee
with $\epsilon$ a characteristic value of $\sigma^{\alpha'}$ and $A_{i_1..i_n}(x')$ are ordinary tensors defined at $x'$.
This is the analogous of an ordinary Taylor expansion in a flat space. Assuming that $T_{\alpha'\beta'}(x,x')$ is known the task is to calculate
 the coefficients  $A_{i_1..i_n}(x')$ of  (\ref{taylor}) . This requires the analysis of the coincident points limit in (\ref{taylor}) and all its derivatives \cite{synge}, \cite{vega}. 
Given an arbitrary bitensor $U(x,x')$ this limit is defined by the formula
$$
[U](x')=\lim_{x\to x'}U(x,x').
$$
In these terms it follows directly that the first coefficient of the expansion (\ref{taylor}) is simply
\be\lb{lc1}
A_{\alpha'\beta'}(x')=[T_{\alpha'\beta'}].
\ee
The calculation of the higher order terms $A_{i_1...i_n}$  in (\ref{taylor}) requires the knowledge of  the coincident point limits of the covariant derivatives $\sigma_{i_1..i_n}$.  These limits can be calculated as follows. First it is true that
\be\lb{cor1}
[\sigma]=0,
\ee
since the distance between to points when $x\to x'$ goes to zero. This condition together with (\ref{bm}) imply that
\be\lb{cor2}
[\sigma_{\alpha}]=0.
\ee
Besides one has
\begin{equation}\lb{cor3}
[\sigma_{\alpha\beta}]=[\sigma_{\alpha'\beta'}]=g_{\alpha\beta}(x'),\qquad [\sigma_{\alpha'\beta}]=[\sigma_{\alpha\beta'}]=-g_{\alpha\beta}(x').
 \end{equation}
 The last relations are intuitive by considering the flat case and can be established by use of the last two formulas together with (\ref{trueee}). Now, In order to calculate the other higher order coincident limits, it is convenient to take two covariant derivatives in (\ref{bm})  to obtain that
\be\lb{dos}
\sigma_{\alpha\beta\gamma}=\sigma^{\delta}_{\alpha\beta\gamma}\sigma_{\delta}+\sigma^{\delta}_{\alpha\beta}\sigma_{\delta\gamma}+
\sigma^{\delta}_{\alpha\gamma}\sigma_{\delta\beta}+\sigma^{\delta}_{\alpha}\sigma_{\delta\beta\gamma}.\ee 
and taking into account (\ref{cor1})-(\ref{cor3}), it follows that
\be\lb{dosdos}
[\sigma_{\alpha\beta\gamma}]=[\sigma^{\delta}_{\alpha\beta}]g_{\gamma'\delta'}(x')+[\sigma^{\delta}_{\alpha\gamma}]g_{\delta'\beta'}(x')+[\sigma_{\delta\beta\gamma}]\delta^{\delta'}_{\alpha'}.\ee
from where it is obtained that
\be\lb{dis}
[\sigma_{\gamma\alpha\beta}]+[\sigma_{\beta\alpha\gamma}]=0.
\ee
Also, since $\sigma$ is a biscalar, it follows that $\sigma_{\alpha\beta}=\sigma_{\beta\alpha}$. By use of this and the Ricci identity
it follows that
$$
2[\sigma_{\alpha\beta\gamma}]=R^\delta_{\alpha\beta\gamma}(x')[\sigma_{\delta}],
$$
This, together with (\ref{cor2}) shows that
\be\lb{cor4}
[\sigma_{\alpha\beta\gamma}]=0.
\ee
Analogously, it can be shown that 
\be\lb{cor44}
[\sigma_{\alpha\beta\gamma'}]=[\sigma_{\alpha\beta'\gamma'}]=[\sigma_{\alpha'\beta'\gamma'}]=0.
\ee
To proceed further requires to take the covariant derivative of (\ref{dos}) to obtain
\be\lb{tres}
\sigma_{\alpha\beta\gamma\delta}=\sigma^{\epsilon}_{\alpha\beta\gamma\delta}\sigma_{\epsilon}+\sigma^{\epsilon}_{\alpha\beta\gamma}
\sigma_{\epsilon\delta}+\sigma^{\epsilon}_{\alpha\beta\delta}\sigma_{\epsilon\gamma}+\sigma^{\epsilon}_{\alpha\gamma\delta}\sigma_{\epsilon\beta}+\sigma^{\epsilon}_{\alpha\beta}\sigma_{\epsilon\gamma\delta}\ee
$$
+\sigma^{\epsilon}_{\alpha\gamma}\sigma_{\epsilon\beta\delta}+\sigma^{\epsilon}_{\alpha\delta}\sigma_{\epsilon\beta\gamma}+\sigma^{\epsilon}_{\alpha}\sigma_{\epsilon\beta\gamma\delta}.
$$
The limit of coincident points in (\ref{tres}) shows that
\be\lb{cordd}
[\sigma_{\alpha\beta\gamma\delta}]+[\sigma_{\alpha\delta\beta\gamma}]+[\sigma_{\alpha\gamma\delta\beta}]=0.
\ee
The last expression can be worked further by taking the derivative of the Ricci identity
$$
\sigma_{\alpha\beta\gamma}=\sigma_{\alpha\gamma\beta}-R^\delta_{\alpha\beta\gamma}\sigma_{\delta},
$$
with respect to $x^\epsilon$ and taking the coincident point limit. The result is
$$
[\sigma_{\alpha\beta\gamma\delta}]=[\sigma_{\alpha\gamma\beta\delta}]+R_{\alpha'\delta'\gamma'\beta'}
$$
Besides, the Ricci identity implies that 
$$
\sigma_{\alpha\beta\gamma\delta}=\sigma_{\alpha\beta\delta\gamma}-R^\epsilon_{\alpha\gamma\delta}\sigma_{\epsilon\beta}-
R^\epsilon_{\beta\gamma\delta}\sigma_{\epsilon\alpha}
$$
and this, together with the symmetry properties of the curvature tensor gives
$$
[\sigma_{\alpha\beta\gamma\delta}]=[\sigma_{\alpha\beta\delta\gamma}].
$$
In these terms it follows that (\ref{cordd}) leads to
\be\lb{cor5}
[\sigma_{\alpha\beta\gamma\delta}]=-\frac{1}{3}(R_{\alpha'\gamma'\beta'\delta'}+R_{\alpha'\delta'\beta'\gamma'}).
\ee
In analogous way the following identities
\be\lb{cor51}
[\sigma_{;\alpha\beta\gamma\delta'}]=\frac{1}{3}(R_{\alpha'\gamma'\beta'\delta'}+R_{\alpha'\delta'\beta'\gamma'}),
\ee
\be\lb{cor52}
[\sigma_{;\alpha\beta\gamma'\delta'}]=-\frac{1}{3}(R_{\alpha'\gamma'\beta'\delta'}+R_{\alpha'\delta'\beta'\gamma'}),
\ee
\be\lb{cor53}
[\sigma_{;\alpha\beta'\gamma'\delta'}]=-\frac{1}{3}(R_{\alpha'\gamma'\beta'\delta'}+R_{\alpha'\delta'\beta'\gamma'}),
\ee
\be\lb{cor54}
[\sigma_{;\alpha'\beta'\gamma'\delta'}]=-\frac{1}{3}(R_{\alpha'\gamma'\beta'\delta'}+R_{\alpha'\delta'\beta'\gamma'}),
\ee
can be proven. 

Once the limits (\ref{cor1})-(\ref{cor54}) are known the coefficients $A_{i_1..i_n}(x')$ of (\ref{taylor}) can be calculated to the third order.
The coincident point limit in (\ref{taylor}) and (\ref{cor1})-(\ref{cor54}) 
give the following recurrence formula
\be\lb{lc2}
A_{\alpha'\beta'}=[T_{\alpha'\beta'}],
\ee
\be\lb{lc3}
A_{\alpha'\beta'\gamma'}=[T_{\alpha'\beta';\gamma'}]-A_{\alpha'\beta';\gamma'},
\ee
\be\lb{lc4}
A_{\alpha'\beta'\gamma'\delta'}=[T_{\alpha'\beta';\gamma'\delta'}]-A_{\alpha'\beta';\gamma'\delta'}-A_{\alpha'\beta'\gamma';\delta'}-
A_{\alpha'\beta'\delta';\gamma'}.
\ee
If $T_{\alpha\beta}$ is known, these formulas allow to determine the expansion coefficients up to order three.

The expansion (\ref{lc2})-(\ref{lc4}) is valid for a bitensor with indices referred to the point $x'$. Consider now the expansion of a bitensor of the form $T_{\alpha'\beta}(x,x')$. In this case one can construct an associated tensor $\widetilde{T}_{\alpha'\beta'}(x,x')$ given by
\be\lb{mentira}
\widetilde{T}_{\alpha'\beta'}(x,x')=g^\beta_{\beta'}(x,x')T_{\alpha'\beta}(x,x'),\ee
which can be expanded by use of  (\ref{lc2})-(\ref{lc4}) and (\ref{taylor}) as
\be\lb{taylor4}
\widetilde{T}_{\alpha'\beta}=B_{\alpha'\beta'}+B_{\alpha'\beta'\gamma'}\sigma^{\gamma'}+
B_{\alpha'\beta'\gamma'\delta'}\sigma^{\gamma'}\sigma^{\delta'}+O(\epsilon^3).
\ee
The formula (\ref{mentira}) can be inverted 
$$ T_{\alpha'\beta}(x,x') =
g_\beta^{\beta'}(x,x')\widetilde{T}_{\alpha'\beta'}(x,x'),
$$
and this together with (\ref{taylor4}) gives that \be\lb{taylor2}
T_{\alpha'\beta}=g^{\beta'}_{\beta}(B_{\alpha'\beta'}+B_{\alpha'\beta'\gamma'}\sigma^{\gamma'}+
B_{\alpha'\beta'\gamma'\delta'}\sigma^{\gamma'}\sigma^{\delta'})+O(\epsilon^3).
\ee
The evaluation of the coefficients $B_{i_1..i_n}$ of this expansion requires the use of  (\ref{cor1})-(\ref{cor54}) and also the coincident point limit of the parallel propagator $g_\alpha^{\beta'}(x,x')$ and its derivative.
These can be calculated as follows. First of all, it is evident from the definition that
$$
[g^{\alpha'}_{\beta}]=\delta^{\alpha'}_{\beta}.
$$
Besides, the parallel transport propagator can be constructed as $g^{\alpha'}_{\beta}=e^{\alpha'}_a e^a_{\beta}$ with $e^a_\beta$ an einbein basis 
which is parallel transported along a geodesic, which means that
$$
e_{a;\beta}^\alpha\sigma^\beta=0.
$$
This implies that
\be\lb{cur1}
g^{\alpha'}_{\beta;\gamma}\sigma^{\gamma}=0.
\ee
A differentiation of the last formula  gives
$$
g^{\alpha'}_{\beta;\gamma\delta}\sigma^{\gamma}+g^{\alpha'}_{\beta;\gamma}\sigma^{\gamma}_\delta=0,
$$
and taking the coincidence limit and using (\ref{cor1})-(\ref{cor3})
give that
\be\lb{cur2}
[g^{\alpha'}_{\beta;\gamma}]=[g^{\alpha'}_{\beta;\gamma'}]=0.
\ee
Further differentiation gives
$$
g^{\alpha'}_{\beta;\gamma\delta\epsilon}\sigma^{\gamma}+
g^{\alpha'}_{\beta;\gamma\delta}\sigma^{\gamma}_\epsilon+g^{\alpha'}_{\beta;\gamma\epsilon}\sigma^{\gamma}_\delta+g^{\alpha'}_{\beta;\gamma}\sigma^{\gamma}_{\delta\epsilon}=0,
$$
The coincident limit of this relation is
$$
[g^{\alpha'}_{\beta;\gamma\delta}]+[g^{\alpha'}_{\beta;\delta\gamma}]=0,
$$
and the use of the Ricci identity gives
$$
2[g^{\alpha'}_{\beta;\gamma\delta}]+R{\alpha'}_{\beta';\delta'\gamma'}=0,
$$
from where it follows that
\be\lb{cur33}
[g^{\alpha}_{\beta';\gamma\delta}]=-\frac{1}{2}R^{\alpha'}_{\beta'\gamma'\delta'}.
\ee
In analogous way it can be shown that
\be\lb{cur3}
[g^{\alpha}_{\beta';\gamma'\delta}]=-\frac{1}{2}R^{\alpha'}_{\beta'\gamma'\delta'},
\ee
\be\lb{cur4}
[g^{\alpha}_{\beta';\gamma\delta'}]=\frac{1}{2}R^{\alpha'}_{\beta'\gamma'\delta'},
\ee
\be\lb{cur5}
[g^{\alpha}_{\beta';\gamma'\delta'}]=\frac{1}{2}R^{\alpha'}_{\beta'\gamma'\delta'},
\ee
With the help of (\ref{cur1}),(\ref{cur5})  together with (\ref{cor1}) and (\ref{cor54}) the limit of coincident (\ref{taylor2}) can be calculated in straightforward manner. 
The result is
\be\lb{lc22}
B_{\alpha'\beta'}=[T_{\alpha'\beta}],
\ee
\be\lb{lc32}
B_{\alpha'\beta'\gamma'}=[T_{\alpha'\beta;\gamma'}]-B_{\alpha'\beta';\gamma'},
\ee
\be\lb{lc42}
B_{\alpha'\beta'\gamma'\delta'}=[T_{\alpha'\beta;\gamma'\delta'}]+\frac{1}{2}B_{\alpha'\epsilon'}R^{\epsilon'}_{\beta';\gamma'\delta'}-B_{\alpha'\beta';\gamma'\delta'}-B_{\alpha'\beta'\gamma';\delta'}-B_{\alpha'\beta'\delta';\gamma'},
\ee
from where the coefficients $B_{i_1..i_n}$  and consequently the expansion (\ref{taylor2}) are determined. 
Finally, in the case that  $T_{\alpha\beta}(x,x')$ is a tensor referred to $x$ one may construct the auxiliary tensor
$$
\widetilde{T}_{\alpha'\beta'}(x,x')=g^{\beta}_{\beta'}g^{\alpha}_{\alpha'}T_{\alpha\beta}(x,x'),
$$ 
and expand it using (\ref{lc2})-(\ref{lc4}) together with (\ref{taylor}) for $\widetilde{T}_{\alpha\beta}(x,x')$ 
\be\lb{taylor3}
T_{\alpha\beta}=g^{\alpha'}_{\alpha}g^{\beta'}_{\beta}(C_{\alpha'\beta'}+C_{\alpha'\beta'\gamma'}\sigma^{\gamma'}+
C_{\alpha'\beta'\gamma'\delta'}\sigma^{\gamma'}\sigma^{\delta'})+O(\epsilon^3),
\ee
with
\be\lb{lc23}
C_{\alpha'\beta'}=[T_{\alpha'\beta}],
\ee
\be\lb{lc33}
C_{\alpha'\beta'\gamma'}=[T_{\alpha'\beta';\gamma'}]-C_{\alpha'\beta';\gamma'},
\ee
\be\lb{lc43}
C_{\alpha'\beta'\gamma'\delta'}=[T_{\alpha'\beta';\gamma'\delta'}]+\frac{1}{2}C_{\alpha'\epsilon'}R^{\epsilon'}_{\beta';\gamma'\delta'}+
\frac{1}{2}C_{\epsilon'\beta'}R^{\epsilon'}_{\alpha';\gamma'\delta'}-C_{\alpha'\beta';\gamma'\delta'}-C_{\alpha'\beta'\gamma';\delta'}-
C_{\alpha'\beta'\delta';\gamma'}.
\ee
these formulas determine the expansion to order two, and can be continued to arbitrary order.

The formulas described above are valid for an arbitrary bitensor $T_{ab}(x,x')$. To give some concrete example consider
for instance $T_{ab}=\sigma_{ab}$. The coincident point limits of this bitensor can be calculated directly from (\ref{cor1})-(\ref{cor54}).
By use of this and the recurrence (\ref{lc2})-(\ref{lc4}) it follows that
\be\lb{ee11}   
\sigma_{ab} = g_a^{\ a'} g_b^{\ b'} \biggl[ g_{a'b'} 
- \frac{1}{3} R_{a'c'b'd'} \sigma^{c'} \sigma^{d'} 
+ \frac{1}{4} R_{a'c'b'd';e'} \sigma^{;c'} \sigma^{d'} \sigma^{e'} 
+  O(\epsilon^4) \biggr],
\ee
\be\lb{e13} 
\sigma_{a'b} = -g^{b'}_{\ b} \biggl[ g_{a'b'} 
+ \frac{1}{6} R_{a'c'b'd'} \sigma^{c'} \sigma^{d'} 
+  O(\epsilon^3) \biggr]
\ee
\be\lb{e12} 
\sigma_{a'b'}=g_{a'b'}-\frac{1}{3}R_{a'c'b'd'}\sigma^{c'}\sigma^{d'}+O(\epsilon^3),
\ee
and also that
\be\lb{e14}
g^a_{b';c} = \frac{1}{2} g^a_{a'} g^{c'}_{c} 
R^{a'}_{\ b'c'd'} \sigma^{d'} + O(\epsilon^2).
\ee
Finally, we quote without proof the expansion for  a bivector $K_a(x,x')$
\be\lb{e13} 
K_a(x,x') = g_a^{a'} \biggl[ K_{a'} - K_{a';c'} \sigma^{c'} 
+ \frac{1}{2} K_{a';c'd'} \sigma^{c'} \sigma^{d'}     
+  O(\epsilon^3) \biggr]. 
\ee 
We turn now our attention on the application of these formulas to the calculation of the singular part of the Green function.

\section{Green function for static geometries in three dimensions}
The divergences of the electrostatic potential $A_t$ at the position of the charge arise due to
fact that the source in the Maxwell equations (\ref{poto}) have a Dirac delta type of singularity. Something analogous happens when a charge
is in front of a perfect conductor, which deforms the field lines and give a net force on the charge. The net force is calculated by subtracting terms 
in the electrostatic field which are divergent at the charge position. The analogous procedure for 
curved geometries was started by Haddamard \cite{haddamard}. This technique is the one employed in our calculation of the charge self -force in a BTZ geometry.
The Taylor like expansions described in the previous section are specially suited for this purpose. 

\subsection{The Haddamard anzatz}
The Maxwell equations (\ref{poto}) for an static charge in a static geometry can be written in the following form
\be\lb{fema}
g_s^{ij}\nabla_i \nabla_j A_t-N^i\partial_i A_t=-2\pi g_{tt} j^t.
\ee
Here $g_s^{ij}(x)$ is the spatial part of the metric and $\nabla_i$ the spatial Levi-Civita connection, both evaluated at the observation position $x$.
In addition
\be\lb{aa}
N_a=\partial_a \log \sqrt{-g_{tt}}.
\ee
By expressing the potential as
 \be\lb{kum}
A_t=-q\sqrt{-g_{tt}(x')} G(x,x'),
\ee
it follows that $G(x,x')$ satisfies the equation
\be\lb{fema}
g_s^{ij}\nabla_i \nabla_j G_t-N^i\partial_i G_t=-2\pi \delta(x,x').
\ee
Here $\delta(x,x')$ represents the spatial Dirac delta in curved space. It is characterized by the property that
$$
\int f(x')\delta(x,x')\sqrt{g_s}dV_s=f(x),
$$
with $f(x')$ an arbitrary test function defined in a neighborhood of $x$. The Haddamard Green function $G(x,x')$ \cite{haddamard}  is a solution of (\ref{fema}) that has the singularity structure enforced by the Dirac delta source, but does not necessarily respect
the boundary conditions of the problem (such as the behavior in the asymptotic region).  Our strategy will not consist in  calculating the electrostatic potential for the BTZ case by the Haddamard method, since it has been already calculated it in (\ref{exp001})-(\ref{exp002}). But the Haddamard calculation will be useful to identify the singular part of this potential. Once this singular part is removed, the resulting renormalized potential will give directly the net self force on the static charge $q$.

The space time dimension will be kept generically in the following, and we denote it as $d=n+2$. If this dimension $n$ is even then the static solution can be expressed as 
\be\lb{impar}
G(x,x')=\frac{1}{n-1}\frac{U(x,x')}{(2\sigma)^{(n-1)/2}},
\ee
while when $n$ is odd one has \cite{haddamard}
\be\lb{par}
G(x,x')=\frac{1}{n-1}\frac{U(x,x')}{(2\sigma)^{(n-1)/2}}+V(x,x')\log\frac{2\sigma}{\lambda}+W(x,x').
\ee
Here $x'$ is the position of the charge singularity. In both cases one has that $U(x,x)=1$. The functional form for $U$, $V$ and $W$ depends on the space time in consideration.
 For $n$ even one can postulate an expansion of the form
$$
U(x,x')=\sum_{p=0}^\infty U_p(x, x')(2\sigma)^p,
$$
which, when inserted into (\ref{fema}) gives the following recurrence formula
\begin{equation} 
\bigl( 2\sigma^{a} \nabla_a - N^a \sigma_{a} + \nabla^2 \sigma 
+ 2p - n - 1 \bigr) U_p = -\frac{2p-n+1}{(n-1)^2} 
\bigl( \nabla^2 - N^a \nabla_a \bigr) U_{p-1},  
\label{eq:Up_recur} 
\end{equation} 
For $n$ odd instead it is postulated that 
\begin{equation} 
U({x},{x'}) = \sum_{p=0}^{\frac{1}{2} (n-3)} U_p({x},{x'}) (2\sigma)^p
\label{eq:U_sigma_odd} 
\end{equation} 
\begin{equation} 
V({x},{x'}) = \sum_{p=0}^\infty V_p({x},{x'}) 
(2\sigma)^p, \qquad 
W({x},{x'}) = \sum_{p=0}^\infty W_p({x},{x'}) 
(2\sigma)^p,   
\label{eq:VW_sigma} 
\end{equation} 
and sustitution into  (\ref{par}) and  (\ref{fema}) gives instead the following recurrence 
\begin{equation} 
\bigl( 2\sigma^{a} \nabla_a 
- N^a \sigma_{a} + \nabla^2 \sigma - 2 \bigr) V_0 = 
-\frac{1}{2(n-1)} \bigl( \nabla^2 - N^a \nabla_a \bigr)
U_{\frac{1}{2}(n-3)}, 
\label{eq:V0_recur} 
\end{equation} 
\begin{equation} 
\bigl( 2\sigma^{a} \nabla_a 
- N^a \sigma_{a} + \nabla^2 \sigma + 2p - 2 \bigr) V_p 
= -\frac{1}{2p} \bigl( \nabla^2 - N^a \nabla_a \bigr) V_{p-1},   
\label{eq:Vp_recur} 
\end{equation} 
together with
$$
\bigl( 2\sigma^{a} \nabla_a - N^a \sigma_{a} + \nabla^2 \sigma 
+ 2p  - 2 \bigr) W_p = -\frac{1}{p} \bigl( 2\sigma^{a} \nabla_a 
- N^a \sigma_{a} + \nabla^2 \sigma + 4p  - 2 \bigr) V_p 
$$
\be
 - \frac{1}{2p} \bigl( \nabla^2 - N^a \nabla_a \bigr) W_{p-1}.  
\label{eq:Wp_recur} 
\ee 
These equations should be supplemented with (\ref{eq:Up_recur}), which also applies to the odd case.
The freedom in choosing the parameter $\lambda$ or $W_0(x,x')$ corresponds to the gauge transformations in the model.
In four dimensions, these type of recurrences where already considered in \cite{DeWitt}.

\subsection{Singular terms of the Green function in 2+1 dimensions}
The recurrence described in the previous subsection has been analyzed in several situations, for instance in the context of black holes in five dimensions \cite{five}.
Nevertheless, to the best of our knowledge,  it was not applied to three dimensional cases. For this reason we made an independent analysis by use of the Synge calculus and the Hadamard anzatz described in previous section.
Our analysis goes as follows. From (\ref{par}) it is inferred that for $2+1$ dimensions, which corresponds to $n=1$, the biscalar $U(x,x')$ can be set to zero redefining $W(x,x')$.
Thus the recurrence (\ref{eq:Wp_recur}) does not play any role in this case. 
The equation (\ref{eq:V0_recur}) for $V_0$ and (\ref{eq:Vp_recur}) for $V_1$ are reduced in this case to \begin{equation} \label{recul1}
\bigl( 2\sigma^{a} \nabla_a - N^a \sigma_{a} + \nabla^2 \sigma 
- 2 \bigr) V_0 = 0. 
\end{equation} 
 \begin{equation} \label{recul2}
\bigl( 2\sigma^{a} \nabla_a - N^a \sigma_{a} + \nabla^2 \sigma)V_1 = -\frac{1}{2}\bigl( \nabla^2 - N^a \nabla_a )V_0 . 
\end{equation} 
The singular part of the Green function we are interested in  is not just the collection of terms which are divergent when $x\to x'$, but also
those whose first derivatives are divergent in that limit, since they give an infinite force.
To solve (\ref{recul1}) one may postulate 
\be\lb{uss}
V_0=1+a_{a'}\sigma^{a'}+\frac{1}{2}a_{a'b'}\sigma^{a'}\sigma^{b'}+\frac{1}{2}a_{a'b'c'}\sigma^{a'}\sigma^{b'}\sigma^{c'}+O(\epsilon^4),
\ee
with $a_{i_1..i_n}$ coefficients to be determined. In the following the attention will be restricted to terms of order three, since they will contain all the singular pieces.
By introducing (\ref{uss}) into (\ref{recul1}) and taking into account the identities deduced in (\ref{ee11})-(\ref{e14}) 
it follows that
\be\lb{f11}
a_{a'}= -\frac{1}{2} N_{a'}, \ee
\be\lb{f12} 
a_{a'b'} = \frac{1}{2} N_{a';b'} + \frac{1}{4} N_{a'} N_{b'} 
+ \frac{1}{6} R_{a'b'}, \ee
\be\lb{f13} 
a_{a'b'c'} = -\frac{1}{2} N_{(a';b'c')} 
-  \frac{3}{4} N_{(a'} N_{b';c')} 
- \frac{1}{8} N_{a'} N_{b'} N_{c'} 
- \frac{1}{4} N_{(a'} R_{b'c')} 
- \frac{1}{4} R_{(a'b';c')}.
\ee 
In deriving this result one has  to take into account the following result
\be\lb{e133} 
N_a = g_a^{a'} \biggl[ N_{a'} - N_{a';c'} \sigma^{c'} 
+ \frac{1}{2} N_{a';c'd'} \sigma^{c'} \sigma^{d'}     
+  O(\epsilon^3) \biggr], 
\ee 
which follows from (\ref{e13}) by identifying $K_a$ with $N_a$.
Consider now (\ref{recul2}). To solve it, it is enough to consider the following terms
\be\lb{uss2}
V_1=b+b_{a'}\sigma^{a'}+O(\epsilon^2).
\ee
By introducing (\ref{uss2}) into (\ref{recul2}), and taking into account (\ref{ee11})-(\ref{e14}), (\ref{uss}) and (\ref{f11})-(\ref{f13})
the following result is obtained
$$
b=\frac{1}{4}a_{a'}^{a'}-\frac{1}{4}N^{a'}a_{a'},
$$
$$
b_{a'}=-\frac{1}{2}b N_{a'}-\frac{1}{8}a^{b'}_{b'}+\frac{1}{12}a_{b'} R^{b'}_{a`}-\frac{1}{8}N^{b'}a_{b'a'}+\frac{1}{8}a_{b'}N^{b'}_{;a'},
$$
which, by (\ref{f11})-(\ref{f13}) can be finally expressed as
\be\lb{g11}
b=-\frac{1}{8}(N_{;a'}^{a'}-\frac{1}{2}N^{a'}N_{a'}+\frac{1}{3}R')
\ee
\be\lb{g12}
b_{a'}=\frac{1}{16}(\nabla^{'2} N_{a'}+N^{b'}_{;b'}N_{a'}-N^{b'}N_{a';b'}+\frac{1}{2}N^{b'} N_{b'}N_{a`}+\frac{1}{3}R' N_{a'}+
\frac{1}{3}R'_{;a'}),
\ee
The equations  (\ref{eq:Wp_recur}) for $W_i$ are not relevant for us, since they do not contain any singularity.
Therefore from (\ref{recul1})-(\ref{recul2}) it follows that the Green function is of the form 
\be\lb{griin}
G(x,x')=\bigg[1+2b\sigma+a_{a'}\sigma^{a'}+2b_{a'}\sigma^{a'}\sigma+\frac{1}{2}a_{a'b'}\sigma^{a'}\sigma^{b'}+
\frac{1}{2}a_{a'b'c'}\sigma^{a'}\sigma^{b'}\sigma^{c'}\bigg]\log\frac{\sigma(x,x')}{\lambda}+...
\ee
with the coefficients given by (\ref{f11})-(\ref{f13}) y (\ref{g11})-
(\ref{g12}).  This expression is valid for any 2+1 dimensional static space time.

The terms in this expansion whose derivatives are divergent when $x\to x'$ are
\be\lb{griin}
G_{sing}(x,x')=\bigg(1-\frac{g_{tt;a'}}{4g_{tt}}\sigma^{a'}\bigg)\log\frac{\sigma(x,x')}{\lambda}+...
\ee
the other terms give no singularities. This formula combined with (\ref{kum}) gives the singular part of the electrostatic potential.
This is the expression we were looking for. We turn our attention in the application of this formula for the electrostatic problem in the BTZ geometry.

\section{Electrostatic self force in BTZ geometries}
The singular terms of the Green function calculated in (\ref{griin}) are generic for any static three dimensional geometry. In the present 
section the Green function formalism will be specialized to the non rotating BTZ geometry, and applies to the calculation of the electrostatic self force of a static charge $q$ in the outer region. 
The differences between the black hole and wormhole case are a theoretical experiment for distinguishing both cases without reaching the throat or the horizon.

In order to solve this task it is necessary to find singular part of the Green function (\ref{griin}) for the non rotating BTZ geometry. At first sight, the formula requires the calculation
of $\sigma(x,x')$ corresponding to the BTZ metric. However symmetry arguments show that charge self force is radial in the BTZ geometry. Thus, it seems reasonable to limit oneself
to the case in which $x$ and $x'$ lie on the same radial line, which can be chosen as $\theta=0$ without loosing generality.
The distance corresponding to this situation was calculated already in (\ref{diston}) for the case $M=1$ and it follows that
$$
r=l\cosh d(r,r_h),
$$
with $d(r,r_h)$ the distance between the point $r$ and the point of the horizon $r_h$ located at the same radial line $\theta=0$. In fact, by introducing the coordinate
$s$ defined by
\be\lb{curde}
r=l \cosh s, \qquad s=\log\bigg(\frac{r}{l}+\sqrt{\frac{r^2}{l^2}-1}\bigg)
\ee
the spatial BTZ metric with $M=1$
$$
g_2=\frac{dr^2}{\frac{r^2}{l^2}-1}+r^2d\theta^2,
$$
is transformed into
\be\lb{spci}
g_2=l^2(ds^2+\cosh^2s d\theta^2),
\ee
from where it follows that the distance between two points on the same radial line is simply \be\lb{radis}d(r_2,r_1)=l(s_2-s_1),\ee
or in radial coordinates
 \be\lb{radis2}d(r, r')=l\log\bigg(\frac{r}{l}+\sqrt{\frac{r^2}{l^2}-1}\bigg)-l\log\bigg(\frac{r'}{l}+\sqrt{\frac{r'^2}{l^2}-1}\bigg),\ee
In these terms the singular part of the Green function for two points on the same radial line $\theta=0$ is given by
$$
G_{sing}(r, r')=\bigg[1-\frac{r}{2l (\frac{r^2}{l^2}-1)^{3/2}}\bigg(\log\bigg(\frac{r}{l}+\sqrt{\frac{r^2}{l^2}-1}\bigg)-\log\bigg(\frac{r'}{l}+\sqrt{\frac{r'^2}{l^2}-1}\bigg)\bigg)\bigg]$$
\be\lb{griin}
\log\bigg(\frac{1}{\lambda}\log\bigg(\frac{r}{l}+\sqrt{\frac{r^2}{l^2}-1}\bigg)-\frac{1}{\lambda}\log\bigg(\frac{r'}{l}+\sqrt{\frac{r'^2}{l^2}-1}\bigg)\bigg)+...
\ee
The regular or renormalized potential for the BTZ black hole is 
\be\lb{runor}
A^a_{ren}(r,r')=A_t(r,r')-A^a_{sing}(r,r'),
\ee
where the singular part 
$$
A^a_{sing}(r,r')=-q\sqrt{\frac{r'^2}{l^2}-1}\; G_{sing}(r, r'),
$$
follows from (\ref{kum}) and (\ref{griin}) and $A_t(r,r')$ is the black hole solution (\ref{exp002}) evaluated at $\theta=0$, namely
\be\lb{xp001}
A^I_t(r,\theta)=-q\log(r_0)+\sum_{n=1}^\infty  \frac{qn^2 k_n(r_0)h_n(r)}{8M},
\ee
for $r<r_0$ and
\be\lb{xp002}
A^{II}_t(r,\theta)=-q\log(r)+\sum_{n=1}^\infty  \frac{q n^2 h_n(r_0)k_n(r)}{8M},
\ee
for $r>r_0$. The self force for the charge $q$ is calculated by
\be\lb{auto1}
F^{r'}=-q F^{(r')}_{(t)}u^{(t)}=-q (e^{(r')})_{r'}F^{r'}_t u^t=-q\sqrt{g_{r'r'}}g^{r'r'} F_{r't}\frac{1}{\sqrt{-g_{tt}}}
=-q\partial_{r'} A_{ren}(r'),
\ee
with $$A_{ren}(r')=\lim_{r\to r'}A_{ren}(r,r').$$
The self force described above work fine in even space time dimensions, and was applied by one of the authors \cite{faltante}
to reproduce correctly the Will-Simth self force in a Schwarzschild black hole \cite{Will80}. However, in odd dimensions
there exist an axiomatic approach to the self force which suggest that such force may pick additional terms \cite{Quinn}.
This procedure is equivalent to the standard one in four dimensions, but not necessarily in even ones. We will denote the self-force
(\ref{auto1}) as the "naive" self force and in the next sections the effect of the additional terms that arise by the methods proposed in \cite{Quinn} will be considered separately.

Although the expressions found above seem to be easy to deal with, there is a problem when limiting oneself to radial lines. The calculation for the self-force involves taking the coincident limit $r\to r'$. 
The complication is that the electrostatic potential $A^a_{sing}(r,r')$ has a singularity in this limit, which is explicitly seen by looking (\ref{griin}). The full potential (\ref{xp001})-(\ref{xp002}) has the same singularity with opposite sign and the combination (\ref{runor}) is free of singularities by construction. But one expression is given as an infinite series and the other is a single term expression. And this is not suitable for solving the $\infty-\infty$ indetermination. One approach may be to sum the series. This can be done in four dimensions, and the resulting potential is a simple algebraic expression \cite{will80}. Unfortunately, we ignore if there exist such summation formulas in three dimensions. Thus, a different approach should be employed. The one to be used below is based  on the following observation: if one is able
to expand the Green function as a Fourier series of the form
\be\lb{desar}
G^s(x,x')=\frac{G_0(r,r')}{2}+\sum_{n=1}^{\infty}G^s_{nc}(r,r')\cos( n \theta)+\sum_{n=0}^{\infty}G^s_{ns}(r,r')\sin( n \theta),
\ee
then  both (\ref{desar}) and (\ref{exp002}) can be combined into a single non divergent series, which can be approximated to an arbitrary order to obtain the approximated self force.  The symmetry argument requires to take the limit $\theta\to 0$ and $r\to r'$ in order to calculate it. However this Fourier expansion can be performed only if we know $\sigma(x,x')$ for  points on the space time with arbitrary $\theta$ values. This function can be calculated explicitly for the BTZ geometry, and the calculation is performed in the appendix. It is given by
$$
\sigma(r,r',\theta,\theta')=\frac{1}{2}d^2(r,r',\theta,\theta')
$$
with
\be\label{distgeod}
d(\mathbf{x_1},\mathbf{x_2)} = l \cosh^{-1} \left[ \frac{r_1 r_2}{r_h^2} \cosh(\theta_2-\theta_1)- \sqrt{\left(\frac{r_1}{r_h}\right)^2-1} \sqrt{\left(\frac{r_2}{r_h}\right)^2-1} \right],
\ee
An important consistency test of (\ref{distgeod}) is to recover the radial distance (\ref{radis}) when $\theta_2\to \theta_1$, that is, when both points lie on the same radial line. For this, it is convenient
to express (\ref{distgeod}) in terms of the coordinates $(s,\theta)$ defined in (\ref{curde}). The result is
\be\label{dist geod en func dist radial}
d(\mathbf{x_1},\mathbf{x_2)} = l \cosh^{-1} \left[ \cosh s_1 \cosh s_2 \cosh(\theta_2-\theta_1) - \sinh s_1 \sinh s_2 \right].
\ee
From the last  expression it follows that when the points are on the same radial line then $\alpha = 0$  and the distance becomes
$$
d(\mathbf{x_1},\mathbf{x_2)} = l \cosh^{-1} \left[ \cosh s_1 \cosh s_2 - \sinh s_1 \sinh s_2 \right] 
= l \cosh^{-1}\left[ \cosh (s_2 - s_1 ) \right]= l (s_2 - s_1).
$$
Thus, the radial distance (\ref{radis}) has been recovered in this limit, as expected. Since the expression (\ref{distgeod}) for the geodesic distance between two arbitrary points in the BTZ geometry is explicit, the calculation of the singular Green function (\ref{griin}) for the electrostatic problem is immediate. Unfortunately, the expression that is obtained is very complicated and we were not able to find a closed expression for the Fourier coefficients of the expansion. But this is not to be discouraged, since there is a numerical trick that can be employed, which will allow us to approximate the real value of the self force to an arbitrary order. We turn our attention to this trick in the next section. 

\subsection{Fourier expansion of  the singular Green function}
In the previous section the singular part of the Green function has been found explicitly, but not its Fourier expansion. The analytic form of this expansion turns out to be out of our computational methods.  In these situations, the following observation may be useful. In some mathematical applications it may be of interest to study a given function $f(x)$ not in the full domain, but in an small interval $(x_0-\epsilon, x_0+\epsilon)$ around a point $x_0$.  For example, this may be because the points around $x_0$ are the ones that considerably  influence the value of some integral. In these situations it may be valid to find a trial function $g(x)$ such that $g(x_0)=f(x_0)$ and such that they values are very close in the mentioned interval. If this new function $g(x)$ has the additional property that its Fourier expansion is simpler, then it may be advantageous to work with it by considering a large number of terms of  the Fourier expansion. For example, one can approximate the function $f(x)$ around $x_0$ with a rectangle function with small width and with $f(x_0)$ as its height. Note that if all terms are considered, both the exact and the approximated Fourier expansions have the same value at $x_0$, under suitable conditions for $f(x)$ and $g(x)$. 

Based on this idea, our goal is to construct an approximated Green $G_a(r,r',\theta,\theta')$ function which coincides with the exact  one when the two points $x$ and $x'$ lay on the same radial line $\theta=\theta'$.
Since the symmetry of the problem implies that the  force is obtained taking the limit $r\to r'$ for points lying on the same radial line, then the fact that this approximated Green function does not coincide with the exact one outside this line is not relevant for the calculation. This approximated Green function is found in terms of an approximated distance $d^a(x,x')$ which coincide with (\ref{radis}) for two points located on the same radial line.

At first sight there are a large number of possible choices for the approximated Green function, and the idea is choose one which is simple to work with. A physically motivated one is given in the appendix for  mass value $M=1$. The approximated geodesic distance that is obtained is 
\be\lb{dist}
 d^a(x, x_0)= l\sqrt{1+\frac{(\theta-\theta_0)^2}{\bigg(\tan^{-1}\tanh\frac{s}{2}-\tan^{-1}\tanh\frac{s_0}{2}\bigg)^2}}(s-s_0), \qquad 0<\arrowvert \theta-\theta_0\arrowvert<\pi,
 \ee
 \be\lb{dist2}
 d^a(x, x_0)=l\sqrt{1+\frac{(2\pi-\theta+\theta_0)^2}{\bigg(\tan^{-1}\tanh\frac{s}{2}-\tan^{-1}\tanh\frac{s_0}{2}\bigg)^2}}(s-s_0), \qquad \pi<\arrowvert \theta-\theta_0\arrowvert<2\pi.
 \ee
 We will employ this expression in the following by choosing  the $\theta$ values in the range $-\pi<\theta<\pi$. By construction (\ref{dist2}) is not strictly true, but tends to (\ref{radis}) when $\theta_1=\theta_2$, and it is even under the interchange $\theta \leftrightarrow-\theta$, which is a property of the true distance. Note in addition, that   the limit
 \be\lb{lomu}
\lim_{s\to s_0} \frac{(s-s_0)^2}{\bigg(\tan^{-1}\tanh\frac{s}{2}-\tan^{-1}\tanh\frac{s_0}{2}\bigg)^2}=\cosh^2(s_0),
 \ee
 implies that the distance (\ref{dist}) is perfectly regular when $\theta\neq \theta_0$. The same argument holds for (\ref{dist2}).

The approximated world function $\sigma^a(x,x')$ can be found directly found by taking the square of the distance element (\ref{dist}). It is convenient to locate the charge at $\theta_0=0$ by simplicity. Then the singular Green function (\ref{griin}) takes the following form \be\lb{gern}
 G^a_{sing}(x,x')=\bigg[f(r, r')\theta^2+g(r,r')\bigg]\log[a(r,r')\theta^2+b(r,r')],
 \ee
with
$$
a(r,r')=\frac{(s-s')^2}{2\bigg(\tan^{-1}\tanh\frac{s}{2}-\tan^{-1}\tanh\frac{s'}{2}\bigg)^2},
$$
$$
b(r,r')=\frac{(s-s')^2}{2},
$$ 
\be\lb{finca}
f(r,r')=-\frac{r}{2} \frac{d}{dr'}\bigg[\frac{(s-s')^2}{(\tan^{-1}\tanh\frac{s}{2}-\tan^{-1}\tanh\frac{s'}{2})^2}\bigg].
\ee
$$
g(r,r')=2-\frac{r}{2} \frac{d}{dr'}(s-s')^2,
$$
Here $s'(r')$ is given by (\ref{curde}). It is convenient to expand this function in the basis $\exp(in \theta)$ as
\be\lb{desarl}
G^a_{sing}(x,x')=\frac{G_0(r,r')}{2}+\sum_{n=1}^{\infty}G^a_{nc}(r,r')\cos( n \theta)+\sum_{n=0}^{\infty}G^a_{ns}(r,r')\sin( n \theta),
\ee
with the radial coefficients given by
$$
G^a_{nc}(r,r')=2\int_{0}^{\pi}G(x,x')\cos(n \theta)d\theta,
$$
and the analogous definition for $G^a_{ms}(x,x')$. Their explicit value of these integrals is
$$
G^a_{nc}(r,r') = \frac{2i}{a n^3} \Psi^c_n \bigg[ Ci \left( n\pi-i n \sqrt{\frac{b}{a}}\right)- Ci \left( n\pi + i n \sqrt{\frac{b}{a}}\right)  $$
$$-  Ci \left(- i n \sqrt{\frac{b}{a}}\right) + Ci \left(i n \sqrt{\frac{b}{a}}\right) \bigg] $$
\be\lb{larsho}
+ \frac{4}{a n^3} \Psi^s_n \bigg[ Si\left(n\pi+i n \sqrt{\frac{b}{a}}\right)-Si\left(-n\pi+i n \sqrt{\frac{b}{a}}\right) \bigg]\ee
$$+ \frac{2}{a \,n^3}\; \bigg[  4\, a \,f \,n \,\pi\,  \cos(n\pi) \bigg( \log(a \pi^2 + b) +1\bigg)\bigg],
$$
where we have introduced
$$
\Psi^c_n = (2af-agn^2+b f n^2) \sinh\left(n\sqrt{\frac{b}{a}}\right) - 2 \sqrt{a b} f n \cosh\left(n\sqrt{\frac{b}{a}}\right),
$$
$$
\Psi^s_n = (2af-agn^2+b f n^2) \cosh\left(n\sqrt{\frac{b}{a}}\right) - 2 \sqrt{a b} f n \sinh\left(n\sqrt{\frac{b}{a}}\right),
$$
by simplicity. For $n=0$ one has 
$$ \hat G^a_0 (r,r') = 2 \log(a \pi^2+b) \left(g\pi+\frac{f\pi^3}{3} \right) + 4 \left( g - \frac{f b}{3a} \right) \left[  \sqrt{\frac{b}{a}} \tan^{-1}\left( \pi \sqrt{\frac{a}{b}} \right) -\pi \right] - 4\frac{f\pi^3}{g}.$$
This is the part we will be interested in if the charge is at $\theta=0$ since $\sin(n\theta)$ vanish at this location.
The notation $Si(x)$ and $Ci(x)$ denote the integral sine and cosine, whose definitions are given by
$$
Si(x)=\int_0^x\frac{\sin t}{t}dt,
$$
$$
Ci(x)=-\int_x^\infty \frac{\cos t}{t}dt.
$$
Taking this and (\ref{kum}) into account it follows that the singular term of the potential $A^a_{sing}$ are given by
\be\lb{vsing}
A^a_{sing}(x,x')=\frac{A_0(r,r')}{2}+\sum_{n=1}^{\infty}A_{nc}(r,r')\cos( n \theta)+\sum_{n=0}^{\infty}A_{ns}(r,r')\sin( n \theta),
\ee
with
$$
A_{ns}(r,r')=-q\sqrt{-g_{t't'}}G_{ns}(r,r')=-q\sqrt{\frac{r'^2}{l^2}-1}\;G^a_{ns}(r,r'),
$$ 
$$
A_{nc}(r,r')=-q\sqrt{-g_{t't'}}G_{nc}(r,r')=-q\sqrt{\frac{r'^2}{l^2}-1}\;G^a_{nc}(r,r'),
$$
and it is completely determined by (\ref{larsho}). 

\subsection{Analysis of the "naive" self-force}

After calculating the singular terms in the electrostatic potential $A_t$, the renormalized potential $A^a_{ren}(r,r')$ is then find as
$$
A^a_{ren}(r,r')=A_t(r,r')-A^a_{sing}(r,r'),
$$
with $A_t(r,r')$ given by (\ref{exp001})-(\ref{exp002}) for the BTZ black hole and by (\ref{solw333}) for the BTZ wormhole, and $A^a_{sing}(r,r')$ is given by (\ref{vsing}).
As shown in (\ref{auto1}) the self force is simply
\be\lb{simply}
F^{r'}=-q\partial_{r'} A_{ren}(r'),
\ee
with $$A_{ren}(r')=\lim_{r\to r'}A_{ren}(r,r').$$ For the black hole it follows from  (\ref{exp002}) that
$$
A_{(bk)ren}^a=-q\log(r')+q\sqrt{\frac{r'^2}{l^2}-1}\;G_{0}(r,r')+q\sum_{n=1}^\infty  \bigg[\frac{ n^2 h_n(r')k_n(r')}{8}+\sqrt{\frac{r'^2}{l^2}-1}\;G_{nc}(r')\bigg],
$$
with
\be\lb{lemoto}
G_{nc}(r')=\lim_{r\to r'}G^a_{nc}(r,r').
\ee
The calculation of (\ref{lemoto}) requires the knowledge of the following limits
$$
\lim_{r\to r'}a(r,r')=2\cosh^2 s'= \frac{2r'^2}{l^2},
$$
$$
\lim_{r\to r'}b(r,r')=0,
$$
$$
\lim_{r\to r'}f(r,r')=-\cosh^2 s'=-\frac{r'^2}{l^2},
$$
$$
\lim_{r\to r'}g(r,r')=2,
$$
which follows directly from the definitions (\ref{finca}). From this it follows that (\ref{lemoto}) is
\be\lb{lemoto22}
G_{nc}(r')=(-1)^n \frac{4\pi}{n^2} \frac{r^2}{r_h^2} \left[   2\log \left(\frac{r}{r_h}  \right) +\log(2\pi^2)+1 \right] + \frac{2}{n}\left( \frac{2}{n^2} \frac{r^2}{r_h^2}-1 \right) Si(n\pi),
\ee
where in this case $r_h=l$. For $n=0$ one has
\be\lb{lemoto220}
G_0(r')=2 \left[2\log \left(\frac{r}{r_h}\right)+\log(2\pi^2) \right ]\left( \pi + \frac{\pi^3}{3} \frac{r^2}{r_h^2} \right) - 4  \pi - \frac{4\pi^3}{9}\frac{r^2}{r_h^2}.
\ee
It is important to remark that from (\ref{lemoto22})-(\ref{lemoto220}) it follows that the Green function at $\theta=0$ is given by
$$
G(r)=2 \bigg[2\log \left( \frac{r}{r_h}  \right)+\log(2\pi^2)\bigg] \left( \pi + \frac{\pi^3}{3} \frac{r^2}{r_h^2} \right) - 4  \pi - \frac{4\pi^3}{9}\frac{r^2}{r_h^2}
$$
$$
+S_1 \frac{r^2}{r_h^2} \left[   2\log \left( \frac{r}{r_h}  \right) +\log(2\pi)+1 \right] + S_2\frac{r^2}{r_h^2}+S_3
$$ 
with the series $S_i$ given by
$$
S_1=4\pi\sum_{n=1}^{\infty}(-1)^n \frac{1}{n^2} ,\qquad S_2=4\sum_{n=1}^{\infty} \frac{1}{n^3},\qquad S_3=2\sum_{n=1}^{\infty} \frac{Si(n\pi)}{n}
$$
The first two are convergent, the third is divergent. This is as expected, since this is the part of the Green function which is singular at coincident point limits. 
We turn now the attention for the self-force corresponding to the two boundary conditions discussed in section 4.

\subsubsection{The self force first type of boundary conditions}

The results discussed above together with the potential (\ref{exp001})-(\ref{exp002}) gives the following renormalized potential for the black hole
$$
A^a_{bh}(r) =  q \sum_{n=1}^\infty  \bigg\{ \frac{ n^2 h_n(r)k_n(r)}{8}+\sqrt{ \frac{r^2}{r_h^2}-1}
\bigg[ (-1)^n \frac{4\pi}{n^2} \frac{r^2}{r_h^2} \left(   \log \left( 2 \pi^2  \frac{r^2}{r_h^2}  \right) +1 \right)
+ \frac{2}{n}\left( \frac{2}{n^2} \frac{r^2}{r_h^2}-1 \right) Si(n\pi) \bigg] \bigg\}
$$
\be\lb{renormolon}
- q \log \left( \frac{r}{r_h} \right)+ q \sqrt{ \frac{r^2}{r_h^2}-1} \; \left[ 2 \log \left( 2 \pi^2  \frac{r^2}{r_h^2}  \right) \left( \pi + \frac{\pi^3}{3} \frac{r^2}{r_h^2} \right) - 4  \pi - \frac{4\pi^3}{9}\frac{r^2}{r_h^2} \right] \ee
The wormhole potential follows from (\ref{solw333}) and the last expression (\ref{renormolon}), the result is
\be\lb{eve}
A^a_{(wh)ren}(r)=A^a_{(bk)ren}(r)-\frac{q}{2}\sum_{n=1}^\infty \bigg[\frac{1}{r_g h_n(r_g)}+\frac{n^2k'_n(r_g)}{8}\bigg]
 \frac{h^2_n(r')}{h'_n(r_g)}.
\ee
These expressions is a combination of two series which are divergent at $r\to r'$, but whose divergent behavior cancel exactly. Therefore when evaluated at the radial line $\theta=\theta'$, both (\ref{renormolon})-(\ref{eve}) are finite and has finite derivatives, which give the self force.

The following figures show the approximate behavior for  the force $F_t$ , for some values of $n$ between $n=15$ and $n=45$ for the black hole case.  This graphs were obtained by use of MATHEMATICA. The graphs show the behavior for large value of $r$, which corresponds to $s\sim 1$. The functional form of the force is always the same in all the orders we considered namely, it starts to grow from the horizon $s=0$ till it attach some maximum value, and then decays to zero at $s=1$ (the asymptotic region). But as larger orders are considered, the maximum start to grow and to move closer to the asymptotic region $s=1$.  This suggest us that when all the series is summed up, this maximum becomes infinitely large, and the line $s=1$ is an asymptote for the function. In other words, the decaying part is just an artifact of the truncation. If this is correct and is not related to a numerical problem in the program we are using, then it follows that the self force is repulsive and grows indefinitely as the charge goes to the asymptotic region.  The repulsive nature is the same as in the Schwarzschild four dimensional counterpart \cite{will80}. But is mandatory to interpret the strange behavior at the infinite.  The explanation for such non typical behavior may be that the boundary conditions considered are not correct, and include extra charges at the asymptotic region. Thus it is convenient to analyze the "right" boundary conditions to see if this behavior is improved.

We have also considered the wormhole case. The effect of the throat  is to give a contribution to the force of opposite sign for the black hole. This effect is also seen in the Schwarzschild case \cite{faltante}. In that case, this contribution change the sign of the force in some region of the space time. However, with the numerical precision we are working with, we did not find such change of the behavior in the BTZ geometry, the net effect of the throat is just a change in the numerical value of  the force. But this is not conclusive, and perhaps this effect may appear by considering higher orders in the series.

\begin{figure}[h!]
\centering
\includegraphics{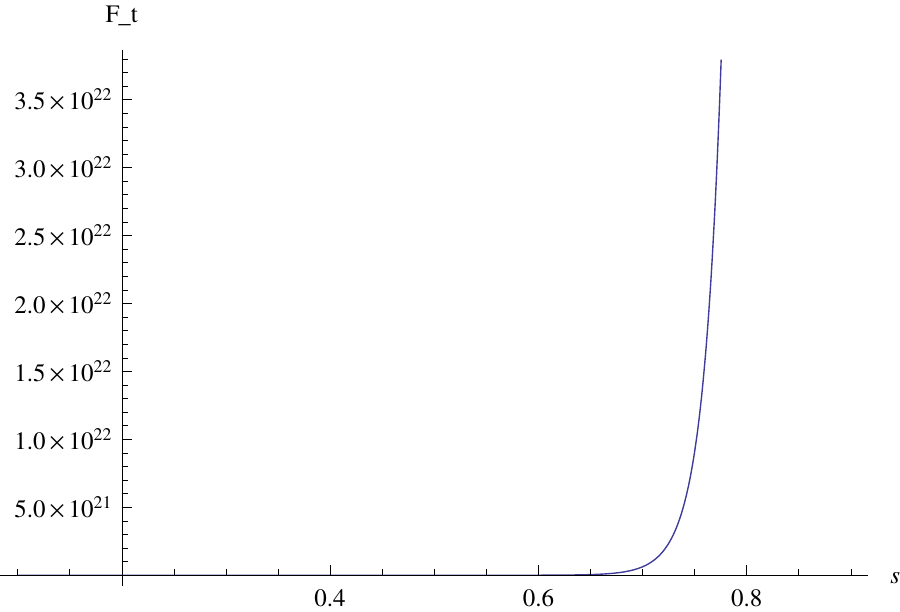}
\caption{The black hole self force function of $s=1-r^{-2}$, from $s=0$ to $s=0.9$ to order $n=25$. The force seems to grow, but the next figure shows that it attach a maximum.}
\label{figuruno1}
\end{figure}

\begin{figure}[h!]
\centering
\includegraphics{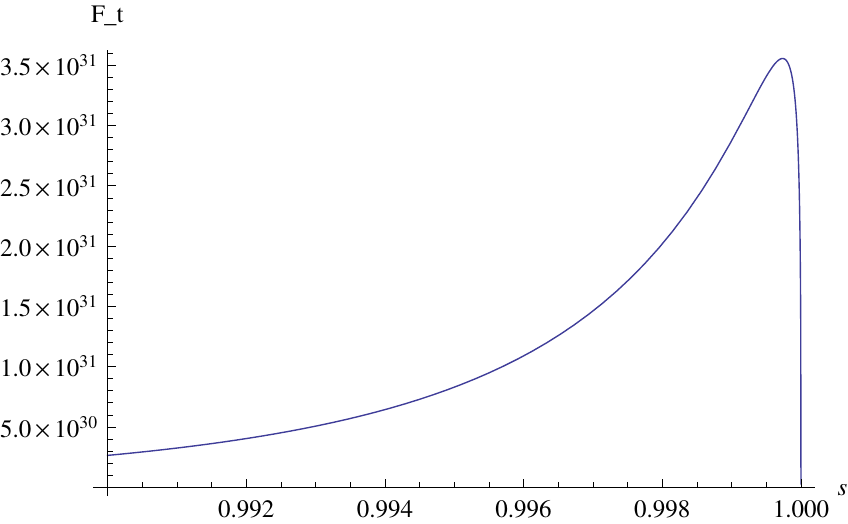}
\caption{The behavior of the self force near the boundary $s=1$. It reaches a maximum and then goes to zero. We argue that this is just an effect of the truncation of the series and that the line $s=1$ is in fact an asymptote.}
\label{figuruno2}
\end{figure}

\begin{figure}[h!]
\centering
\includegraphics{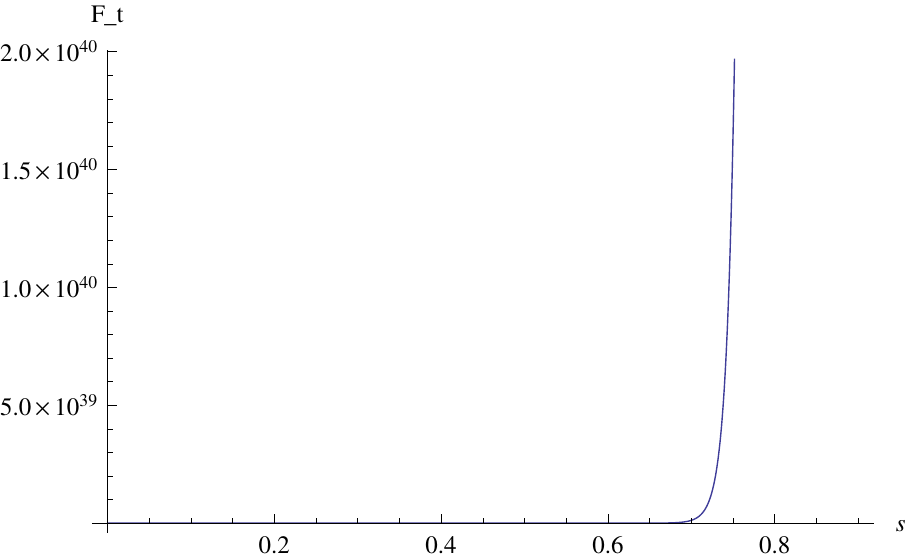}
\caption{The force for $n=45$  in the interval from $s=0$ to $s=0.9$.}
\label{figuruno5}
\end{figure}

\begin{figure}[h!]
\centering
\includegraphics{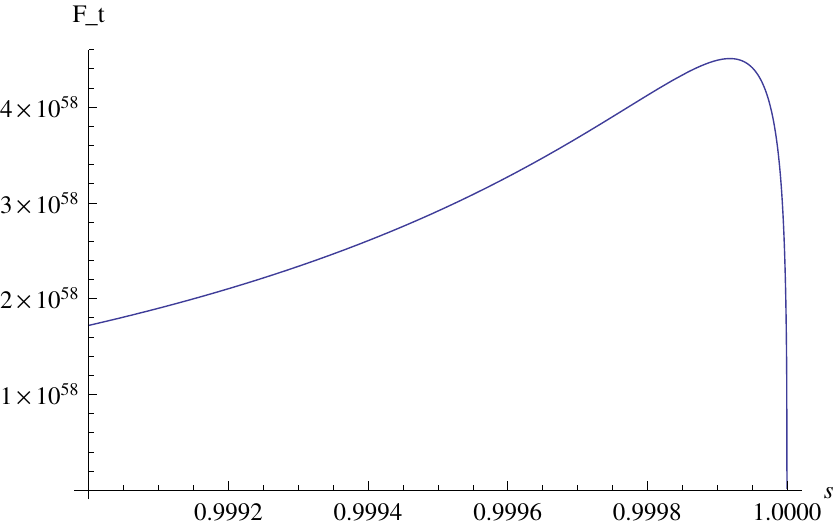}
\caption{The behavior near of the horizon $s=1$ with $n=45$. The qualitative form of the curve is similar to the case $n=25$ but the maximum is again more pronounced and closer to $s=1$.
This suggest that when all the series is summed, the maximum will tend to infinite and the vertical line $s=1$ will be the asymptote. See below for an explanation of this seemingly strange result.}
\label{figuruno6}
\end{figure}

\subsubsection{The self force corresponding to the right type of boundary conditions}
For this type of boundary conditions, the potential follows from (\ref{pu002}) instead of (\ref{exp002}). The result is
$$
A^a_{bh}(r) =  \frac{q}{2}\sum_{n=1}^\infty \bigg\{ \Gamma(1+\frac{in}{2}) \Gamma(1-\frac{in}{2}) h_n(r)f_n(r)+\sqrt{ \frac{r^2}{r_h^2}-1}
\bigg[ (-1)^n \frac{4\pi}{n^2} \frac{r^2}{r_h^2} \left(   \log \left( 2 \pi^2  \frac{r^2}{r_h^2}  \right) +1 \right)
$$
\be\lb{renormolonww}
+ \frac{2}{n}\left( \frac{2}{n^2} \frac{r^2}{r_h^2}-1 \right) Si(n\pi) \bigg] \bigg\}- q \log \left( \frac{r}{r_h} \right)
+q \sqrt{ \frac{r^2}{r_h^2}-1} \; \left[ 2 \log \left( 2 \pi^2  \frac{r^2}{r_h^2}  \right) \left( \pi + \frac{\pi^3}{3} \frac{r^2}{r_h^2} \right) - 4  \pi - \frac{4\pi^3}{9}\frac{r^2}{r_h^2} \right].
\ee

\begin{figure}[h!]
\centering
\includegraphics{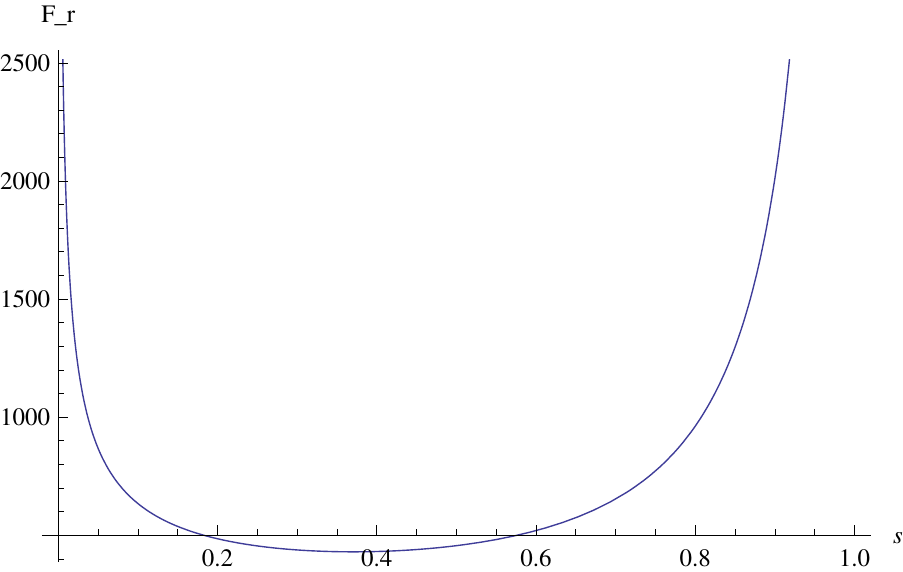}
\caption{The black hole self force function of $s=1-r^{-2}$, from $s=0$ to $s=0.9$ to order $n=20$. The force seems to be positive and divergent near the horizon and at the asymptotic boundary, and there is an intermediate values of $s$ where it is negative.}
\label{figuruno7}
\end{figure}

\begin{figure}[h!]
\centering
\includegraphics{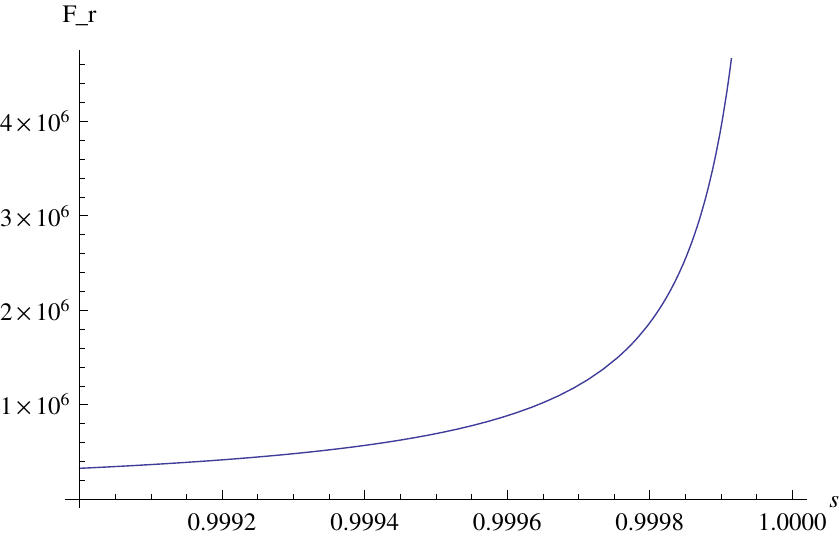}
\caption{The behavior of the self force near the boundary $s=1$, which shows that it is divergent.}
\label{figuruno8}
\end{figure}

\begin{figure}[h!]
\centering
\includegraphics{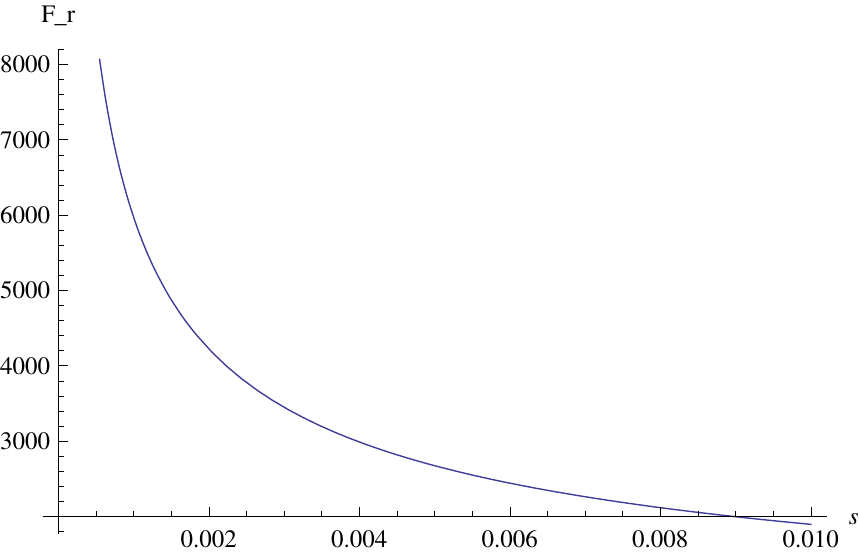}
\caption{The behavior of the self force near the horizon $s=0$, which shows that it is also divergent.}
\label{figuruno9}
\end{figure}

\begin{figure}[h!]
\centering
\includegraphics{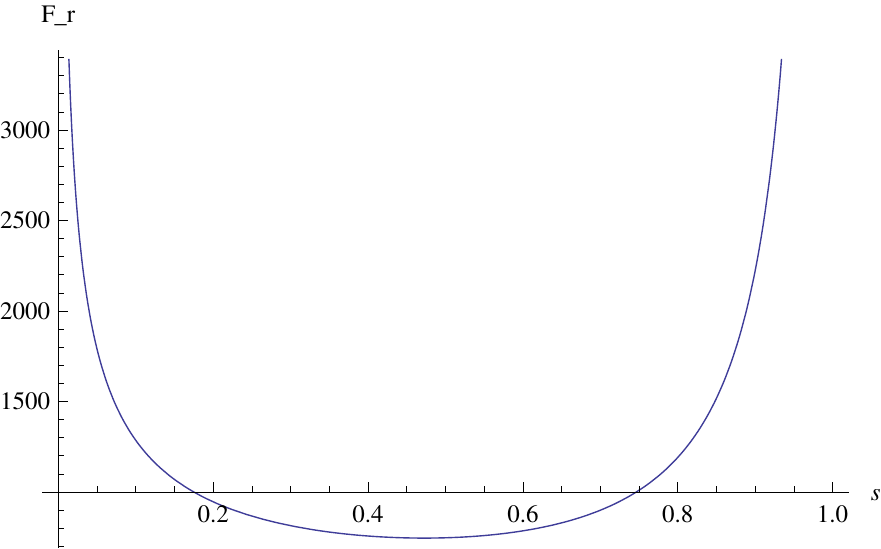}
\caption{The black hole self force function of $s=1-r^{-2}$, from $s=0$ to $s=0.9$ to order $n=40$. The result is qualitative the same as for $n=20$.}
\label{figuruno10}
\end{figure}

\begin{figure}[h!]
\centering
\includegraphics{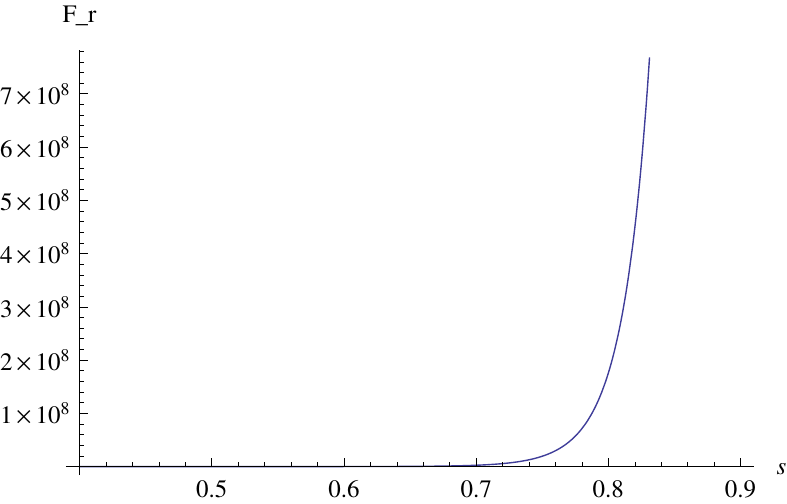}
\caption{The wormhole self force when the throat is $r_g=3$, with $s$ varying from $s=0$ to $s=0.9$. The effect of the throat is to increase the slope of the asymptote. It is not clearly seen in this graph if there is a region where the force is negative.}
\label{figuruno11}
\end{figure}

\begin{figure}[h!]
\centering
\includegraphics{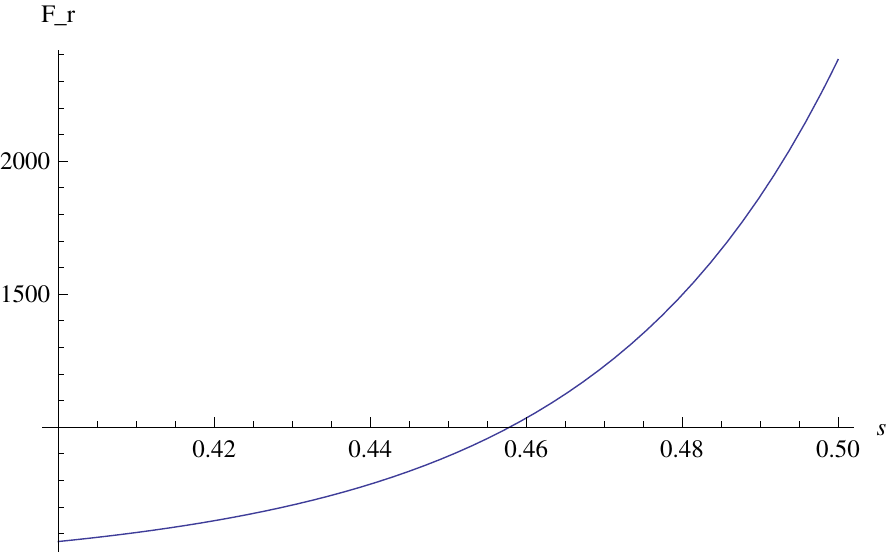}
\caption{The wormhole self force when the throat is $r_g=3$, with $s$ varying from $s=0.4$ to $s=0.5$. Here it is seen that there is a negative force at some region. The difference with the black hole is that the point where the sign change is closer. So the location of that turning point may be a clue for distinguishing the geometries. }
\label{figuruno12}
\end{figure}

The wormhole potential follows from (\ref{solp333}) and the last expression (\ref{renormolon}), the result is
\begin{equation}\label{eveww}
A^a_{(wh)ren}(r)=A^a_{(bk)ren}(r)-\frac{q}{2}\sum_{n=1}^\infty \bigg[\frac{ h_n(r)}{r_g h_n(r_g)h'_n(r_g)}
+\Gamma(1+\frac{in}{2}) \Gamma(1-\frac{in}{2})\frac{h_n(r)k'_n(r_g)}{2 h'_n(r_g)}\bigg] h_n(r)
\end{equation}
The figures (\ref{figuruno7})-(\ref{figuruno10}) show the self force behavior for this case. It is surprising to see that the self force is again divergent at the asymptotic boundary and also at the horizon. From a formal point of view, this divergent behavior can be seen directly by looking (\ref{renormolonww}). The logarithmic terms are multiplied for functions of $r$ and there are terms proportional to $r^2$ as well in this expression. The derivatives with respect to $r$ of all these terms are not bounded for large $r$. This may seem the cause for the growth of the force at large values. On the other hand, th terms proportional $h_n(r)f_n(r)$ go fast to zero since, as discussed in previous sections the derivative of  $f_n(r)$ goes as $1/r^3$ and the one of $h_n(r)$ goes as $1/r$.  Thus these terms does not influence what happens at large $r$. But they arguably the cause of the divergence near the horizon. As also discussed in previous section the function $f_n(r)$ possess the fastest decaying conditions. However, the behavior of $f_n(r)$ at the horizon is highly non regular and, even though $h_n(r)\to 0$, the full contribution is not bounded. In other words, the fastest decaying conditions also give the worse behavior at the horizon. 

The other surprising behavior is that the self force is repulsive near the horizon and far away, but at intermediate radial positions, it is attractive. This is in contrast with the Schwarzschild case in four dimensions, where it is always repulsive \cite{will80}. However, without a proper understanding to the origin of the divergences, this analysis does not make any sense. We turn to this point below.

\section{Is the force calculated the real one?}
The procedure of previous section computes a regular Green function. However, there may exist other regular Green functions as well, since the addition to a solution of the homogeneous equation
is still a full solution of the problem.  Therefore the statement that the calculated self-force is the real one must be taken with care. To give an example, the half-advanced-half-retarded
Green function, and the regularized ("R") force of the Detweiler-Whiting 
decomposition \cite{DW} are regular at the location of the particle. Nevertheless only the 
"R" force corresponds to the correct expression for the
self-force. 

The method presented in the previous section works fine in even space time dimensions. In fact, this method was applied by one of the authors in \cite{faltante} to reproduce correctly the Will-Smith self force for a Schwarzschild black hole in four dimensions \cite{will80}. However, the BTZ geometry is odd dimensional and this case is more tricky, as suggested by the axiomatic approach presented in \cite{Quinn} and applied recently for five dimensional space times in \cite{five}. This approach introduce extra terms in our calculation. These terms vanish in four dimensions and do not spoil the results obtained in \cite{faltante}. But in odd dimensions, this may not be the case. A further motivation for studying these terms is that they may cure the divergent behavior at the asymptotic boundary found in the previous section. However, it will be shown below that this is not the case.

The self force calculated in the previous sections was given by
$$
F_a=\partial_a A-\partial_a A_{sing},
$$
with $A$ the full potential satisfying the specific boundary conditions of the problem and $A_{sing}$ the part of the potential whose values and the values of its derivatives
$\partial_a A_{sing}$ are divergent when the position $x$ of the observation point tends to the charge position $x'$.
The approach of \cite{Quinn} states instead that the self force should be calculated as
\be\lb{qw}
F_a=\partial_a A-\partial_a A_{sing}+<\partial_a A_{sing}>,
\ee
 where $<\partial_a A_{sing}>$ is the average of the quantity inside over an small surface witth $s=cte$ in the limit $s\to 0$, from which all the contributions that are divergent in the limit $s\to 0$ are removed. As it will be shown below, this average will pick some additional quantities in the calculation. The derivative is $\partial_a$ is taken on a point of that surface. Now, since the singular part is given by
$$
A_{sing}=-q\sqrt{\frac{r'^2}{l^2}-1}G_{sing}(x,x'),
$$
and the derivatives are taken at $x$, it follows that the relevant derivative to be taken is 
$$
\partial_a G_{sing}(x,x')=h^{a'}_a\bigg[\bigg(1-\frac{g_{tt;b}}{4g_{tt}}\sigma^{b}\bigg)\frac{\sigma_{a'}}{\sigma}-\frac{g_{tt;a'}}{4g_{tt}}\log\frac{\sigma(x,x')}{\lambda}+...\bigg],
$$
up to a factor which is evaluated at the charge position.  In order to make the average, it is convenient to use Riemann normal coordinates $x^a$. This procedure is extensively reviewed in \cite{vega} and we quote only the main ingredients. These coordinates are characterized in terms of the Synge world function as follows. The derivative $\sigma^{a'}=-x^a$ and 
$$
2\sigma=-2 \delta_{ab}x^a x^b=s^2,
$$
is the squared proper distance from the point $x'$. The parallel propagator in these coordinates is approximated by
$$
h_b^{a'}=\delta_a^b-\frac{1}{6}R^{a'}_{b'c'd'}x^b x^c+O(s^3).
$$
By denoting $x^a=s\Omega^a$ with $\Omega^a$ some angular variables it follows that
$$
\partial_a G_{sing}(x,x')= \frac{2\Omega_{a}}{s}- \frac{g_{tt;b}}{2g_{tt}}\Omega^a\Omega^{b}-\frac{g_{tt;a'}}{4g_{tt}}\log\frac{s^2}{\lambda}+...
$$
In particular, one has that $\Omega_a= \delta_{ab}\Omega^b$. The average along the surface $s=const$ is defined by
$$
<...>=\frac{\int(....)dA}{\int dA},
$$
with $dA$ the area element of the surface. If this surface is parameterized by polar angles $\theta_a$ then this and the relation $x^a=s \Omega^a$ implies that the metric is
$$
ds^2=s^2 h_{ab}\Omega^a_i \Omega^b_j d\theta^i d\theta^j,
$$
where
$$
\Omega^a_i=\frac{\partial \Omega^a}{\partial \theta_i}.
$$
The area element is then calculated as 
\be\lb{aro}
dA=s^2\bigg(1-\frac{s^2}{6}R_{a'b'}\Omega^a \Omega^b -\frac{s^3}{12}R_{a'b';c'}\Omega^a \Omega^b \Omega^c+O(s^4)\bigg)d\Omega_n.
\ee
Here $d\Omega_n$ is the area element for the sphere $S^n$, which satisfies
$$
\Omega_n=\int _{S^n} d\Omega_n=\frac{2\pi^{(n+2)/2}}{\Gamma(\frac{n+1}{2})}.
$$
The following identities take place \cite{vega}
$$
\frac{1}{\Omega_n}\int \Omega^a d\Omega_n=0,
$$
$$
\frac{1}{\Omega_n}\int \Omega^a \Omega^b d\Omega_n=\frac{\delta^{ab}}{n+1},
$$
By use of these identities and the definition of the area element (\ref{aro}) it follows that
$$
<\Omega^a>=-\frac{s^3}{6(n+1)(n+3)}\nabla^{a'}R+O(s^4)
$$
\be\lb{udoj}
<\Omega^a\Omega^b>=-\frac{1}{(n+1)}\bigg(\delta^{ab}-\frac{s^2}{6(n+3)}R^{a'b'}+\frac{s^2}{3(n+1)(n+3)}R \delta^{a'b'}\bigg)+O(s^4)
\ee
Then the expression of the averaged singular Green function is
$$
<\partial_a G_{sing}(x,x')>= \frac{2<\Omega_{a}>}{s}- \frac{g_{tt;b}}{2g_{tt}}<\Omega^a\Omega^{b}>-\frac{g_{tt;a'}}{4g_{tt}}\log\frac{s^2}{\lambda}+...
$$
By taking into account the identities (\ref{udoj}) it follows that in the limit $s\to 0$ the surviving terms are 
\be\lb{dull}
<\partial_a G_{sing}>(x')=- \frac{g_{tt;a'}}{6g_{tt}}-\frac{g_{tt;a'}}{4g_{tt}}\log\frac{s^2}{\lambda}+...
\ee
Thus, as in five dimensions \cite{five}, this average picks an extra term which is divergent as $\log s$.  This term can not be absorbed by a redefinition of the parameters
of the problem. The nature of this divergence is different than the one of the previous section, it suggest that in odd dimensions the calculating is sensible to the size of the particle. 
The limit $s\to0$ gives a global divergence of the self force, while the divergence of the previous section is just asymptotic. 

We would like to remark that we have considered the effect of the first term in (\ref{dull}) but it is not enough for avoiding the bad behavior at the infinite.
Thus another interpretation is required. We turn to this point below.

\section{Interpretation of the divergences}

The most striking point of the results presented above is the behavior of the charge self force at the asymptotic region and at the horizon. For the black hole solution, it grows indefinitely when approaching both regions.
For the wormhole, it diverges at the asymptotic boundary.

The physical origin of these divergencies is not easy to visualize. At first sight, it is not strange that the force does not vanish at the asymptotic region. In fact, when the geometry is not asymptotically flat, this situation may happen, as shown for four dimensions in \cite{Kuchar}. When $M=1$, on dimensional grounds, it is expected that the asymptotic self force should be proportional to 
$$
F_{\infty}\sim \frac{e^2}{l}.
$$
Now, the problem of the divergence  at $r\to \infty$ may come from the expression (\ref{dist})-(\ref{dist2}) for the distance, namely
$$
 d^a(x, x_0)= l\sqrt{1+\frac{(\theta-\theta_0)^2}{\bigg(\tan^{-1}\tanh\frac{s}{2}-\tan^{-1}\tanh\frac{s_0}{2}\bigg)^2}}(s-s_0), \qquad 0<\arrowvert \theta-\theta_0\arrowvert<\pi,
 $$
 and the analogous formula for negative angle values. Recall that this distance is a fake one, but it reduces to the true radial distance when $\theta=\theta_0$. The expressions that have been obtained for the self force
 are based on this fake distance. However, since the radial limit is correct, the resulting expressions coincide with the true self force when all the series is summed up. The choice of the fake distance was for simplicity, since the Fourier expansion for the potential is non tractable when the exact distance is considered (the true distance is derived in the Appendix). Nevertheless, an inspection of the formula $d_a(x,x')$ shows the following potential numerical problem. This formula, as shown in the appendix, is similar to the real one when the points $x$ and $x_0$ are almost on the same circle $s\sim s_0$. And coincides with real one when $\theta=\theta_0$.  The distance when the points are at different circles $s<< s_0$ or $s>>s_0$ is more complicated, and it turns out that the expressions differ considerably. Denote these distances as $d_a^c(x,x')$ and $d^r_a(x,x')$ respectively. Both distances are the same when evaluated on the line $\theta=\theta_0$, but when $\theta=\theta_0+\epsilon$ they may differ considerably. Moreover, their difference is a function of the radial coordinate $|d_a^r(x,x')-d^c_a(x,x')|=f(r, \epsilon)$.  The series expansion presented above is convergent to the real solution but, due to the mentioned dependence in $r$, the convergence may be highly non uniform. In other words, one has that
$$
|A_{reg}^n(r)-A_{reg}(r)|< \epsilon \qquad \longleftrightarrow \qquad n> n_0(r),
$$
with $A^n_{reg}$ corresponding to a truncation of the series of $A_{reg}$ to order $n$. This means that,  depending on the values of $r$, lower or higher orders may be required. We interpret that the divergence at $r\to \infty$ arises due to the fact that for large $r$ values higher orders are required, and it is then an artifact of the truncation. 

Another fact that suggest that this numerical problem is the cause of the divergence is the fact that when $r$ and $n$ takes large values the slope of the asymptote seem to grow indefinitely and it moves to the right of the graph ($s\to1$). Arguably when $n\to \infty$ this asymptote moves to $r\to \infty$ ($s=1$) with infinite slope. Note that the point $s=1$, since it corresponds to $r\to\infty$, is not a point in the manifold. Thus, we argue that when the series is summed up, the divergence is an artifact and the limit $F_r$ when $r\to \infty$ is well defined.  We were unable to overcome the problems described above, since the real distance leads to a Fourier expansion which is beyond our calculation technology. 

It may be mentioned that the use of the approximate distance may be useful if a powerful summation formula were available allowing a closed analytical expression for the full potential $A$. This is the situation in four dimensions, as explained in \cite{faltante} and references therein. But, if this formula do exist, we ignore it.

There is a further point to be discussed. When $r>>lM$, then one may consider $M\sim 0$. For this situation, it follows by dimensional analysis that
$$
F_r=\frac{e^2}{l} f(\frac{r}{l}).
$$
If the function $f(x)$ is bounded, then when $l\to \infty$ the self force goes to zero. This is the situation in four dimensional anti De Sitter spaces  \cite{Kuchar}. In this limit the geometry is flat and the self force vanishes.
However, in three dimensions, such limit does not corresponds to a flat geometry, instead for $M=1$ and $J=0$ the  BTZ metric (\ref{bitiz}) reduce to 
$$
g_3=-dr^2 +dt^2+r^2d\theta^2,
$$
in the limit $l\to\infty$. Note that $r$ now is a "time" coordinate and the compact radius is "r-time" dependent. Thus the derivatives $\partial_r A$ do not make sense as a self force, and perhaps this divergence is not signaling anything wrong.  However,  our opinion is that the behavior of the infinite is more likely a numerical artifact. This is suggested by the divergence at the horizon, for which we have no explanation. Thus, it is arguably a numerical problem what is causing the divergent behavior.  Our results are of course non conclusive, and further research is mandatory to elucidate the real behavior at $r>>l$.

\section{Summary}
In the present work the particularities of the electrostatics of point charge in front of a BTZ black hole and wormhole were considered. This geometry is not flat neither asymptotically flat, and this results in several unexpected features. For instance, every radial functions for the electrostatic problems goes to zero at the asymptotic horizon. This complicates the boundary condition analysis, since there is no a clear criteria of which radial function discarding at the region between the charge position and the asymptotic infinite. We argue that, although both radial functions are well behaved, the right choice of the boundary condition is to choose the radial function which has the fastest decay when the radial coordinate is large. This situation is the one which imitates better the electrostatic in flat space or in asymptotically flat spaces.

After discussing the boundary condition for the problem, we followed the program 
\cite{eoc12} -\cite{lipatova} in order to calculate the self force of a charge in front of the BTZ black hole horizon and compare it to the corresponding wormhole geometry.
In addition we have considered the axiomatic approach of \cite{Quinn}.
Unfortunately, both program seems to be harder to achieve. The regularized electrostatic potential is presented as a Fourier expansion, and we had no available summation formula
for obtaining a closed analytical formula for such potential. As a consequence, we were forced to use an approximated geodesic distance which reduces to the real one when the two comparison 
points are on the same radial line. The use of such distance allows a series expansion which converges to the real vector potential, but arguably the convergence is highly non uniform. This is reflected in that
the obtained diagrams for the self force, when truncated to a given order, have an asymptotic divergence. We interpret as fictitious, arising for the fact that a large number of orders are needed for large radial values. 
However, it is not clear which behavior should be expected in this geometry at asymptotic values, since it is not asymptotic flat neither it admits a flat limit. 

There is a further possibility however. The boundary conditions that we have chosen are the ones that decay faster at the asymptotic region. This condition imitates ordinary electrostatics and is the one that is more comfortable to our intuition. However, there may be some symmetries of the BTZ space time similar to those in higher dimensional black holes \cite{zelnikov}. We do not know how to implement this symmetry in the BTZ geometry and how is reflected in the Green function, but perhaps the result  does not respect this boundary condition. After all, what the present work is showing is that flat geometry based intuitions may lead to wrong results. In any case our results are not conclusive about this issue. A further investigation about this point may be of special interest for the future.
 \\

{\bf Acknowledgements:}  The authors are supported by CONICET (Argentina).

\section*{Appendix }  
\renewcommand{\theequation}{A.\arabic{equation}}

\setcounter{equation}{0}  

\subsection*{Exact BTZ geodesic distance}
Our next task is to find the geodesic distance $d(r,r',\theta,\theta')$ between two arbitrary points of the space time. This distance should reduce to (\ref{radis}) or (\ref{radis2})  when $\theta'\to \theta$.
The geodesic curve joining two points $x$ and $x'$ on the BTZ geometry is describe by the Euler-Lagrangian equations derived with the following action 
\be\lb{acdic}
d(\mathbf{x_1},\mathbf{x_2}) = \int_{\mathbf{x_1}}^{\mathbf{x_2}} {\sqrt{\frac{l^2}{r^2-{r_h}^2} \left( \frac{dr}{d\lambda} \right)^2 + r^2 \left( \frac{d\theta}{d\lambda} \right)^2 } d\lambda},
\ee
The geodesic parameter can be chosen as  $\lambda = \theta$. The corresponding  lagrangian is
$$
\mathcal{L}(\dot r, r) = \sqrt{\frac{l^2 \dot r^2}{r^2-{r_h}^2}  + r^2}.
$$
The value of  (\ref{acdic}) evaluated at the geodesic curve is by definition the geodesic distance. 
Since $\mathcal{L}$ does not depend explicitly on the "time" parameter $\lambda=\theta$ there exists a conserved quantity, the hamiltonian
\be\label{cdmov}
H = \frac{\partial\mathcal{L}}{\partial \dot r} \dot r - \mathcal{L} = \frac{-r^2}{\sqrt{\frac{l^2 \dot r^2}{r^2-{r_h}^2}  + r^2}}.
\ee
In terms of this constant, the lagrangian evaluated in the physical trajectory is
$$
\mathcal{L}^2 = g_{rr} \dot r^2 + g_{\theta\theta}^{-1} H^2 \mathcal{L}^2,
$$
from where it follows that
\be\label{lagrangiano integrable}
\mathcal{L}^2 = g_{rr} g_{\theta\theta} \dot r^2 \left(g_{\theta\theta}- H^2 \right)^{-1}.
\ee
In these terms the geodesic distance is given as
$$
d(\mathbf{x_1},\mathbf{x_2}) = \int_{\lambda_1}^{\lambda_2} \sqrt{\frac{g_{rr}g_{\theta\theta}}{\left(g_{\theta\theta}- H^2 \right)}} \; \dot r \, d\lambda
$$
and since $\dot r d\lambda = dr$, then by taking into account the explicit expression for the metric tensor, it follows that
\be\label{distancia en func de L}
d(\mathbf{x_1},\mathbf{x_2}) = l \int_{r_1}^{r_2} \frac{r \; dr}{\sqrt{\left(r^2 - {r_h}^2 \right)  \left(r^2 - H^2 \right)}} = l \log \left[\frac{ \sqrt{ {r_2}^2 - {r_h}^2 } + \sqrt{ {r_2}^2 - {H}^2 }}{\sqrt{ {r_1}^2 - {r_h}^2 } + \sqrt{ {r_1}^2 - {H}^2 }} \right] .
\ee
The distance (\ref{distancia en func de L}) is not completely determined, unless the value of $H$ is given in terms of the two positions $x$ and $x'$.
This is achieved as follows. From (\ref{cdmov})  and 
 (\ref{lagrangiano integrable}) it is deduced that
$$
\left( \frac{dr}{d\theta} \right)^2 = \frac{\mathcal{L}^2}{g_{rr} g_{\theta\theta}} \left(g_{\theta\theta}- H^2 \right) \left(\frac{g_{\theta\theta}}{H \mathcal{L}} \right)^2 
= \frac{r^2}{H^2 l^2} \left( r^2 - {H}^2\right) \left(r^2-{r_h}^2 \right),
$$
and since $\dot r^2 > 0 $ and $ r>r_h$, then  $H^2 < r^2 $. From the last equation it follows that
$$
\theta_2 - \theta_1 = \pm l H \int_{r_1}^{r_2} \frac{dr}{r \sqrt{r^2- H^2 } \sqrt{r^2 -{r_h}^2}}.
$$
The last integration is elementary, the result is
\be\label{angulo en funcion de L}
\theta_2 - \theta_1 =  \log \left[ \frac{r_1}{r_2} \left(\frac{r_h \sqrt{{r_2}^2- H^2 } + L \sqrt{{r_2}^2 -{r_h}^2}}{r_h \sqrt{{r_1}^2- H^2 } + L \sqrt{{r_1}^2 -{r_h}^2}} \right)\right].
\ee
The expression just obtained can be worked out further by defining 
$$e^\alpha= e^{\theta_2 - \theta_1} = \frac{ \sqrt{H^{-2}- {r_2}^{-2} } + \sqrt{{r_h}^{-2} -{r_2}^{-2}}}{\sqrt{H^{-2}- {r_1}^{-2} } + \sqrt{{r_h}^{-2} -{r_1}^{-2}}}.$$
This definition is equivalent to
$$e^{\alpha/2} \sqrt{H^{-2}- {r_1}^{-2} } - e^{-\alpha/2} \sqrt{H^{-2}- {r_2}^{-2} } = e^{-\alpha/2}  \sqrt{{r_h}^{-2} - {r_2}^{-2}} - e^{\alpha/2} \sqrt{{r_h}^{-2} -{r_1}^{-2}}.,$$
and by taking the square of both members it follows after some algebra that \be\lb{eq con tildes}
\tilde\Lambda = \tilde\eta + (H^{-2}- {r_h}^{-2}) \cosh\alpha,
\ee
where the following quantities  $$\tilde\Lambda = \sqrt{H^{-2}- {r_2}^{-2} }\sqrt{H^{-2}- {r_1}^{-2} },\qquad \tilde\eta = \sqrt{{r_h}^{-2} - {r_1}^{-2}} \sqrt{{r_h}^{-2} -{r_2}^{-2}},
$$ 
has been introduced. On the other hand (\ref{distancia en func de L}) implies that
$$e^{d/l} = \frac{ \sqrt{ {r_2}^2 - {r_h}^2 } + \sqrt{ {r_2}^2 - {H}^2 }}{\sqrt{ {r_1}^2 - {r_h}^2 } + \sqrt{ {r_1}^2 - {H}^2 }},$$
and  procedure analogous to the one made above shows that
\be \label{eq sin tilde}
\Lambda = \eta + ({r_h}^{2}- H^2) \cosh\frac{d}{l}.
\ee
where in this case
$$
\Lambda = \sqrt{ {r_1}^2 - {H}^2 }\sqrt{ {r_2}^2 - {H}^2 },
\qquad
 \eta =\sqrt{ {r_1}^2 - {r_h}^2 }\sqrt{ {r_2}^2 - {r_h}^2 }. 
 $$
By noticing the relations
 $$ \eta= r_1 r_2 r_h^2 \tilde\eta,\qquad \Lambda = r_1 r_2 H^2 \tilde\Lambda,$$ 
then comparison between (\ref{eq con tildes}) and (\ref{eq sin tilde}) shows that
$$
{r_h}^{2} \cosh\frac{d}{l}  =  r_1 r_2 \cosh\alpha - \eta.
$$
From the last expression the explicit geodesic distance $d$ can be obtained, the result is
$$
d(\mathbf{x_1},\mathbf{x_2)} = l \cosh^{-1} \left[ \frac{r_1 r_2}{r_h^2} \cosh(\theta_2-\theta_1)- \sqrt{\left(\frac{r_1}{r_h}\right)^2-1} \sqrt{\left(\frac{r_2}{r_h}\right)^2-1} \right],
$$
This is the formula (\ref{distgeod}) obtained in the text.
\subsection*{Approximate geodesic distance}
In the following the attention will be restricted to BTZ geometries with mass values $M=1$. Based on the arguments given above and from the fact that the calculation of the self force requires to take the limit $\theta\to 0$, our next task is to find an approximated distance function $d^a(x,x')$ which is equal to $d(x,x')$ when $\theta\to 0$ and satisfy the mandatory periodicity conditions. The upper index $a$ enforce the fact that the distance is not exact and this notation will be used repeatedly in the following. At first sight there 
is a variety of candidates for $d^a(x,x')$. But one that is physically motivated is the following. Consider the spatial BTZ  metric (\ref{spci})  in the coordinates $(s,\theta)$. The geodesic distance is
\be\lb{judi}
d=l \int\sqrt{s'^2+\cosh^2 s}\;d\theta,
\ee
with $s'$ the derivate of $s(\theta)$ with respect to $\theta$. The conserved quantity $H$ related to the $\theta$ independence on the lagrangian is expressed in these coordinates as
$$
H=\frac{l \cosh^2 s}{\sqrt{s'^2+\cosh^2 s}}.
$$
It is important for the following to discuss the physical significance of this quantity. When the geodesic line is close to a circle, then $s'\sim 0$ and $H$ is essentially the radius of the circle $r=l \cosh s$.
Instead, when the geodesic  is close to a radial line, then $s'\to\infty$ and $H\to 0$ in this situation. The definition of $H$
 implies that \be\lb{sub}
s'^2=\frac{l^2\cosh^4 s}{H^2}-\cosh^2s,
\ee
and for nearly radial lines one can neglect the second term in (\ref{sub}) since $H<<1$, thus concluding that
$$
s'^2\sim \frac{l^2 \cosh^4 s}{H^2}.
$$
For nearly circular geodesics one has that 
$$
H^2\sim (l^2-\epsilon^2) \cosh^2 s,
$$
with $\epsilon^2<<l^2$, and within this approximation (\ref{sub}) becomes
\be\lb{arm}
s'^2\sim \frac{\epsilon^2}{2l^2} \cosh^2s.
\ee
We have explicitly checked that the equation  (\ref{arm}) leave to the simplest expression for $d^a(x,x')$.  The equation (\ref{arm}) implies that
\be\lb{geo2}
 \frac{1}{c}\frac{ds}{d\theta}=\pm \cosh(s),
 \ee
 $c$ being a constant related to $\epsilon^2$ whose value is fixed by the initial conditions of the problem. 
 The equation (\ref{geo2}) can be integrated in elementary form to give
 $$
 \theta-\theta_0=\pm \frac{1}{c }\bigg[\tan^{-1}\tanh\frac{s}{2}-\tan^{-1}\tanh\frac{s_0}{2}\bigg].
 $$
This shows that the constant $c$ is given in terms of the initial and final positions by
 \be\lb{geo3}
 c=\frac{\bigg[\tan^{-1}\tanh\frac{s}{2}-\tan^{-1}\tanh\frac{s_0}{2}\bigg]}{(\theta-\theta_0)}.
 \ee
 Inserting (\ref{geo2}) and (\ref{geo3}) in (\ref{judi}) give a result the following expression for the approximated geodesic distance 
$$
 d^a(x, x_0)= l\sqrt{1+\frac{(\theta-\theta_0)^2}{\bigg(\tan^{-1}\tanh\frac{s}{2}-\tan^{-1}\tanh\frac{s_0}{2}\bigg)^2}}(s-s_0), \qquad 0<\arrowvert \theta-\theta_0\arrowvert<\pi,
 $$
 $$
 d^a(x, x_0)=l\sqrt{1+\frac{(2\pi-\theta+\theta_0)^2}{\bigg(\tan^{-1}\tanh\frac{s}{2}-\tan^{-1}\tanh\frac{s_0}{2}\bigg)^2}}(s-s_0), \qquad \pi<\arrowvert \theta-\theta_0\arrowvert<2\pi.
 $$
 This is the expression (\ref{dist})-(\ref{dist2}) found in the text.

\end{document}